\documentclass[%
 reprint,
superscriptaddress,
preprintnumbers,
bibnotes,
amsmath,amssymb,
aps,
prl,
showkeys 
]{revtex4-1}

\usepackage{amsmath}
\usepackage{graphicx}

\begin{document}

\title{A Rigorous Method of Calculating Exfoliation Energies from First Principles}

\author{Jong Hyun Jung}
\affiliation{Department of Physics and Astronomy, Seoul National University, Seoul 08826, Korea}

\author{Cheol-Hwan Park}
\email{cheolhwan@snu.ac.kr}
\affiliation{Department of Physics and Astronomy, Seoul National University, Seoul 08826, Korea}

\author{Jisoon Ihm}
\email{jihm1@postech.ac.kr}
\affiliation{Department of Physics, Pohang University of Science and Technology, Pohang 37673, Korea}

\date{\today}

\begin{abstract}

The exfoliation energy, the energy required to peel off an atomic layer from the surface of a bulk material, is of fundamental importance in the science and engineering of two-dimensional materials. 
Traditionally, the exfoliation energy of a material has been obtained from first principles by calculating the difference in the ground-state energy between (i) a slab of $N$ atomic layers ($N \gg 1$) and (ii) a slab of $N-1$ atomic layers plus an atomic layer separated from the slab.  In this paper, we
prove that the exfoliation energy can be obtained {\it exactly} as the difference in the ground-state energy between a bulk material (per atomic layer) and a single isolated 
layer.
The proposed method is (i) tremendously lower in computational cost than the traditional approach since it does not require calculations on thick slabs, (ii) still valid even if there is a surface reconstruction of any kind, (iii) capable of taking into account the relaxation of the single exfoliated layer (both in-plane lattice parameters and atomic positions), and (iv) easily combined with all kinds of many-body computational methods. As a proof of principles, we calculated exfoliation energies of graphene, hexagonal boron nitride, MoS$_2$ and phosphorene using density-functional theory.  In addition, we found that the in-plane relaxation of an exfoliated layer accounts for 5\% of one-layer exfoliation energy of phosphorene while it is negligible ($<\,0.4\%$) in the other cases. 

\end{abstract}


\maketitle

Since the exfoliation of the first two-dimensional (2D) material, graphene \cite{Novoselov2004}, experimentalists have tried to separate more and more 2D materials composed of one or few atomic layers from mechanical exfoliation \cite{Novoselov2016,Ferrari2015}.  Certainly, a very important criterion for the feasibility of mechanical exfoliation is its energy cost. Computation of this energy cost is very important not only in explaining why certain materials are exfoliated easily but also in predicting which 2D materials can be separated from bulk compounds as a guide to experimentalists \cite{Mounet2018,Peng2016,Lebegue2010}. In some studies, the exfoliation energy has been
defined as
the energy cost of peeling the top layer from a surface of a bulk crystal \cite{Wang2015,Gould2013,Liu2012,Spanu2009,Ziambaras2007,Hanke2011,Bjorkman2012,Chen2013}. Alternatively, in a completely different context,
the exfoliation energy has also been defined as the
interlayer binding energy of a layered material, i.\,e.\,,
the energy (per layer) required
to separate all the layers of the bulk \cite{Schabel1992,Zacharia2004,Ortmann2006,Hasegawa2007,Ashton2017}.
Here, we adopt the first definition since it is directly related to the actual mechanical exfoliation. The exfoliation energy defined in this way is a measure of
the difficulty in peeling a layer from the surface \cite{Gould2013,Ziambaras2007,Chen2013,Bjorkman2012,Hanke2011}.

Previously, in line with its first definition,
the exfoliation energy was calculated as the difference between the total energy
of a thick slab composed 
of several ($N$) atomic layers and
that of an isolated atomic layer plus the remaining ($N-1$)-layer
slab \cite{Gould2013,Ziambaras2007,Jing2017,Zhao2014,Li2016,Jiao2015}. In these
calculations, periodic boundary conditions were used, and
the isolated layer and the remaining slab were placed in the same
supercell (with a large vacuum region in between). This method neglects a possible change in the
lattice parameters of the exfoliated layer.
Furthermore, in order to obtain a convergent result in simulating the  surface of a bulk crystal, a large $N$ was required, which resulted in heavy computational load.  We call this method a ``slab method.''

In order to reduce the computational load, some previous studies approximated the exfoliation energy
(according to the first definition \cite{Wang2015,Gould2013,Liu2012,Spanu2009,Ziambaras2007,Hanke2011,Bjorkman2012,Chen2013},
i.\,e.\,,
the energy cost of exfoliation from a surface)
by the interlayer binding energy.  However, interlayer binding energy is a rather conceptual quantity and it cannot be measured directly from an experiment. Although the closest process relevant to interlayer binding energy would be a complete dissolution of a layered material into individual layers in a liquid, it should involve unwanted interactions between the solvent molecules and the layered material.
These studies assumed that the interlayer interaction could be treated as pairwise additive, that the relaxation of in-plane lattice parameters of an exfoliated layer is negligible, and that the layers near the surface have the same atomic structure as those deep inside the bulk so that the surface relaxation or reconstruction can be neglected.  Importantly, even if there are some limited cases in the phase space of the electronic-structure methods, energy functionals, and materials where this approximation is shown to be good, there has been no general proof that the interlayer binding energy should be a good approximation to the exfoliation energy.  Shulenburger et al.\ \cite{Shulenburger2015} regarded
the value of the exfoliation energy as somewhere in between the values of the interlayer binding energy of the bulk material and the binding energy of the two-atomic-layer system, i.e., the energy required to separate the two atomic layers attached to each other.  Sch\"utz et al.\ \cite{Schutz2017} divided the exfoliation energy of phosphorene into the binding energy of an isolated two-atomic-layer system and the remainder, and they approximately obtained this remainder by comparing the total energies of the two- and three-atomic-layer systems obtained from coupled cluster calculations.

In this paper, we propose a conceptually rigorous method to calculate the exfoliation energy
from first principles at a much lower computational cost than the slab method.
We first prove that the exfoliation energy
is {\it rigorously} the same as the difference between the ground-state energy (per layer) of the bulk and that of an isolated layer. Now, the calculation using the usual periodic boundary 
conditions is very simple because it requires only the primitive unit cell of a bulk crystal
and the supercell containing a single atomic layer.
Since this method does not involve any thick slab calculations, a huge reduction in
the computational cost occurs. 
Moreover, our method naturally takes into account any possible surface reconstructions
and the relaxation of both the atomic positions near the surface
and the in-plane lattice parameters of the peeled layer. Lastly, our method can be naturally combined with any kind of many-body techniques.

By straightforwardly extending this method,
we also show below how to inexpensively calculate the energy required to exfoliate $n$ layers at a time from the surface of a bulk crystal, which we call
the $n$-layer exfoliation energy ($n=1,\,2,\,3,\,...$).
As an application of our method, we calculate the $n$-layer exfoliation energies
of graphene, hexagonal boron nitride (hBN), MoS$_2$, and phosphorene
from first principles using the density-functional theory.

\begin{figure}[tbp]  
    \centering
    \includegraphics[height=9cm]{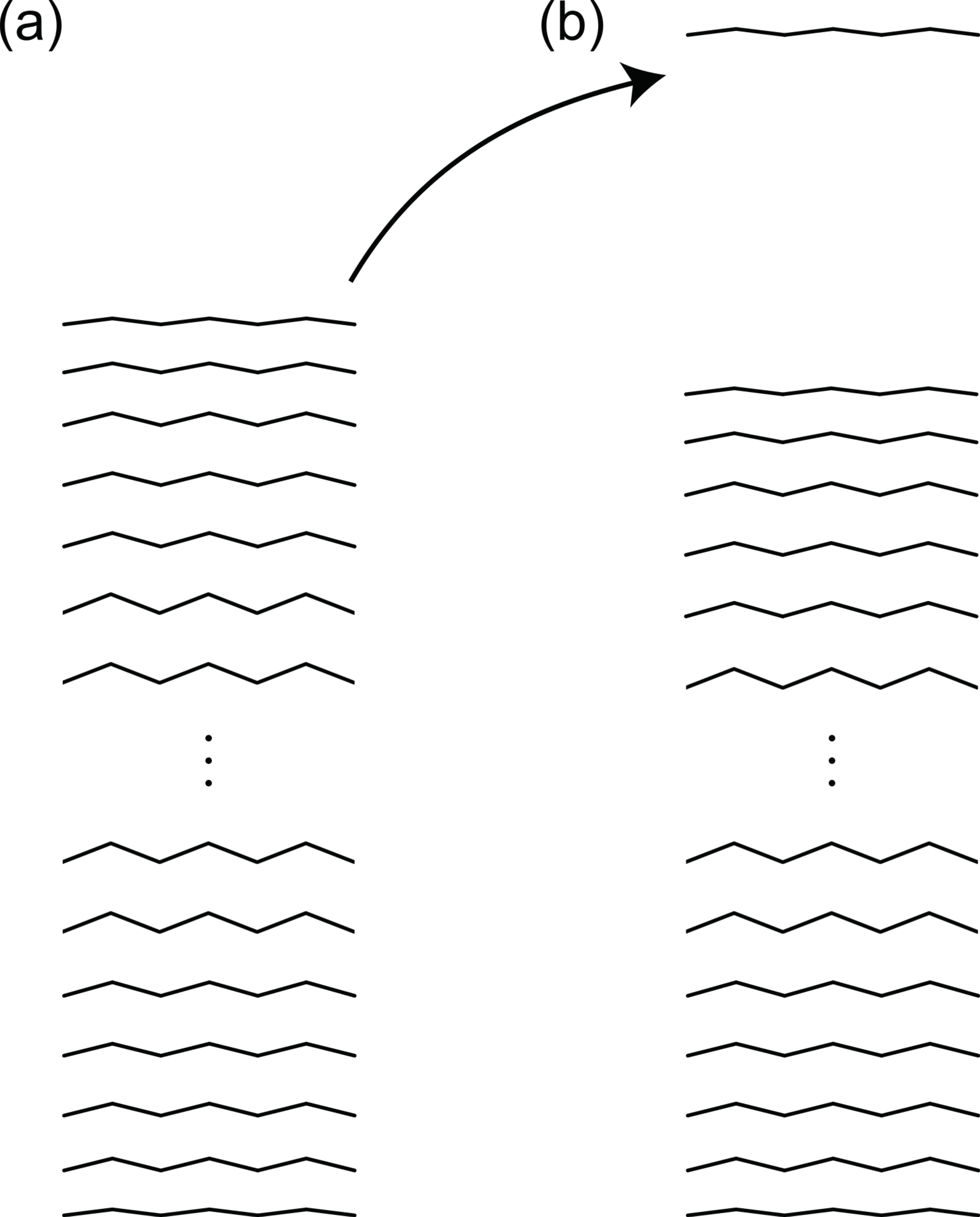}
    \caption{\label{figure-previous} 
    The traditional slab method for calculating the exfoliation energy. It is obtained from the difference in the ground-state energies of two configurations: (a) a slab in vacuum consisting of many atomic layers and (b) the topmost layer of the slab being separated (see arrow) and the atomic positions of both the single layer and the remaining slab being relaxed.
    }
\end{figure}

Figure~\ref{figure-previous} illustrates
the traditional slab method of calculating the exfoliation energy.
The exfoliation energy is obtained from the
energy difference before and after
separating an atomic layer from the slab
consisting of many atomic layers.
The total energies of the two configurations shown in
Figs.~\ref{figure-previous}(a) and~\ref{figure-previous}(b) need to be
calculated from first principles. In such calculations, the slab should
contain a large number ($N$) of layers
to simulate the surface of a bulk crystal, and the energy convergence for increasing $N$ should be checked as well.
For this reason, the slab method is computationally heavy.

\begin{figure*}[tbp]  
    \centering
    \includegraphics[height=16cm]{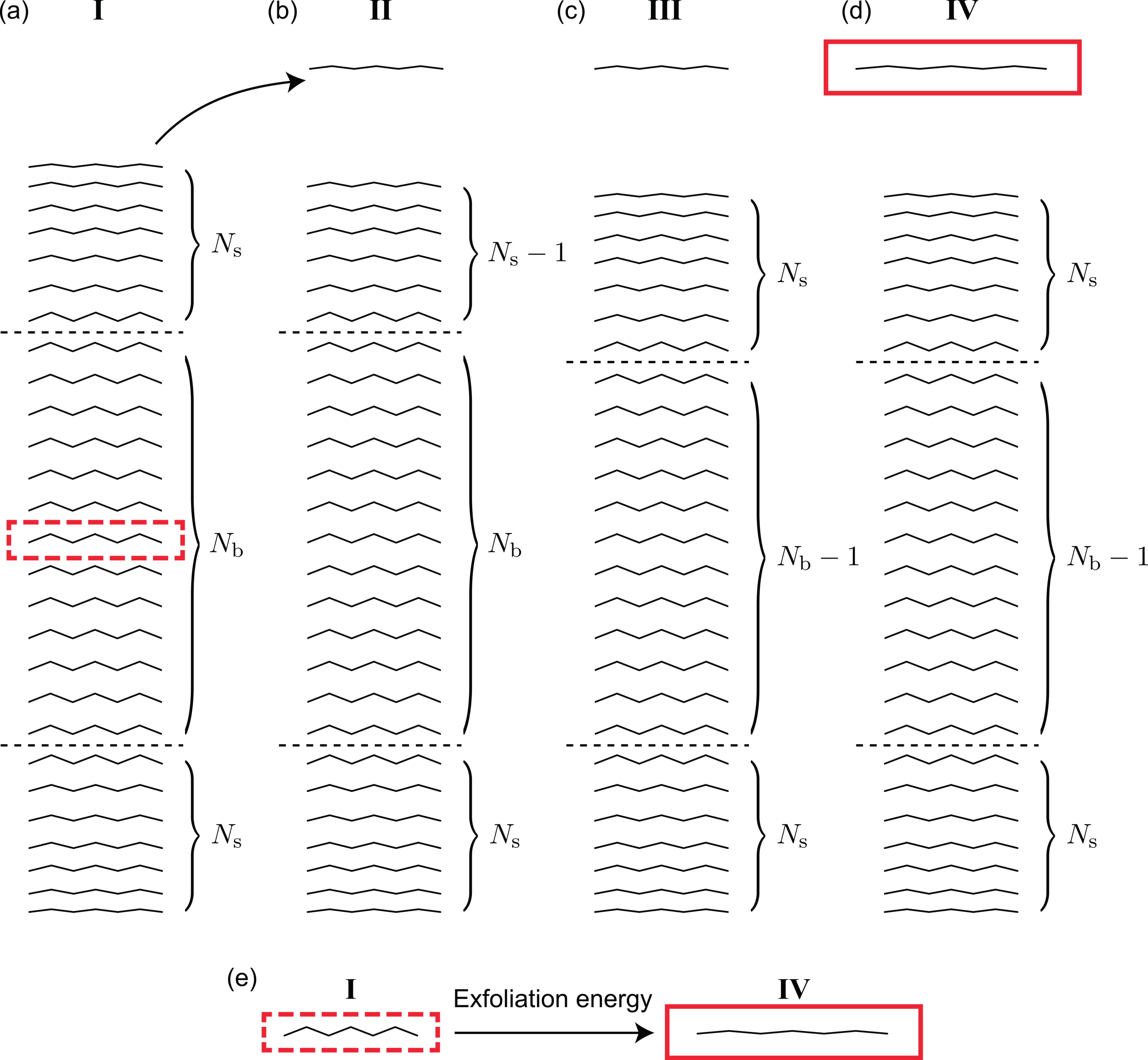}
    \caption{\label{figure-new} 
A schematic of the proposed method for calculating the exfoliation energy. (a)-(d) The conceptual steps for the exfoliation process (see text). The horizontal dashed lines represent the (hypothetical) bulk-surface boundaries.
(e) The exfoliation energy is the energy difference between configurations I and IV, which reduces to the difference in the ground-state energies per layer of the bulk and the fully relaxed atomic layer in vacuum.
    }
\end{figure*}

On the other hand, Fig.~\ref{figure-new} explains
our proposed method of calculating the exfoliation energy,
by dividing the exfoliation process into four consecutive configurations for clarity in presentation. 
We confine our discussion here to materials whose basis atoms of the bulk primitive unit cell
can be contained within one atomic layer (Fig.~\ref{figure-new}).
Later we will generalize our discussion 
to most 2D materials (actually to all known 2D materials
that have been mechanically exfoliated from bulk crystals). 
Initially, before the exfoliation (Configuration I in
Fig.~\ref{figure-new}(a)), 
the crystal has two (upper and lower) surfaces and the surface reconstructions (to form a superstructure) as well as atomic relaxations may exist on the surface regions. The atomic relaxations may occur down to $N_{\mathrm{s}}$ layers from the outermost layer. Our argument below is valid even in the case of very large $N_{\mathrm{s}}$. $N_{\mathrm{b}}$ is the number of bulk layers not influenced by the existence of the surface. 

Now assume that the topmost atomic layer is
separated from the slab (Fig.~\ref{figure-new}(b)). In this step (I$\to$II),
the entire topmost layer is simply translated far away in the surface-normal direction. All the atomic positions of the remaining crystal ($N_{\mathrm{b}} + 2 N_{\mathrm{s}} - 1$ layers) are also fixed (frozen) for the moment.
Then, we allow the atomic positions of the remaining slab to rearrange to the lowest-energy configuration while
those of the removed layer are still held fixed (Fig.~\ref{figure-new}(c)).
In this step (II$\to$III),
the number of layers in the upper surface is recovered to
$N_\mathrm{s}$ and accordingly the number of atomic layers in the bulk is
reduced from $N_\mathrm{b}$ to $N_\mathrm{b}-1$.
Finally, the separated layer in vacuum is allowed to have the lowest-energy atomic configuration 
(Fig.~\ref{figure-new}(d)). In this step (III$\to$IV), the exfoliated
layer may expand or compress in the in-plane directions, and thus the corresponding lattice constants may change.
Other changes may take place as well. For example, atomic buckling in the monolayer plane may occur or the surface reconstruction of the layer which was present before exfoliation may disappear. 
The exfoliation energy is
the change in the total energy of the system in going from
configuration I to IV (Fig.~\ref{figure-new}(e)).
Configuration IV has one less layer in the bulk part than
configuration I, while the two configurations have exactly the same
surfaces. 
Therefore, the energy required to exfoliate
a layer equals the difference in the energy of a layer in bulk and
that of a separate layer in vacuum in their respectively
relaxed geometries.
This equality is not approximate but exact.

Although the logic of our proposed method for calculating exfoliation
energies based on the above configurations (I--IV) was straightforward,
the result is rather unexpected.
What we have shown is that the exfoliation energy is not at all  dependent on
the change in atomic positions (including reconstructions) in the surface region
as the topmost atomic layer is removed.
The reason for the simplicity of our method is that all these complications
cancel out exactly and do not appear in the difference between configurations I and IV.

We provide a mathematical proof to complement the previous pictorial explanation on the equality of exfoliation energy $E_{\mathrm{e}}$  ($>0$) and interlayer binding energy $E_{\mathrm{b}}$  ($>0$).  We will prove an equivalent statement: $ \forall  \epsilon > 0, \ |E_{\mathrm{e}}-E_{\mathrm{b}} |< \epsilon $.  Let $T_N$ be the total energy of a slab composed of $N$ unit layers in its relaxed geometry with possible reconstructions or superstructures.  (Each unit layer may be composed of more than one atomic layer.  For example, a single layer of MoS$_2$ contains three atomic layers.)  Since the area of an in-plane unit cell depends on $N$, we define the energies $E_{\mathrm{e}}$, $E_{\mathrm{b}}$, and $T_N$ as quantities per bulk in-plane unit cell.   
Obviously, the energy $E_0 \equiv  \lim_{N \to \infty} T_N / N $ in the thermodynamic limit is nothing but the well-defined total energy of the bulk crystal per (bulk) unit cell.  Then, $E_{\mathrm{b}}$ can be written as $E_{\mathrm{b}} = \lim_{N \to \infty} (T_1-T_N/N) =  T_1 - E_0.$  Likewise,  $E_{\mathrm{e}}$  can be written as  $E_{\mathrm{e}} = \lim _{N \to \infty}(⁡T_1+T_{N-1}-T_N) $ and it also is a physically well-defined quantity.  We thus have the following relations.
\begin{eqnarray}
\label{eq:Nb}
\forall \epsilon >0, \, \exists N_{\mathrm{b}} (\epsilon)  \textrm{ such that if } N \geq N_{\mathrm{b}} (\epsilon) \nonumber\\
\textrm{ then } |E_{\mathrm{b}}-(T_1-T_N/N)|<\epsilon/4. \\
\label{eq:Ne}
 \forall \epsilon>0, \, \exists N_\mathrm{e} (\epsilon) \textrm{ such that if } N \geq N_\mathrm{e} (\epsilon) \nonumber\\
 \textrm{ then } |E_\mathrm{e}-(T_1+T_{N-1}-T_N )|<\epsilon/4.
\end{eqnarray}
We define $N_\mathrm{c} \equiv \max⁡[N_\mathrm{e} (\epsilon),N_\mathrm{b} (\epsilon)]$ for a given $\epsilon$.
If we apply Eq.~\ref{eq:Ne} for $N=2N_\mathrm{c}, N =2N_\mathrm{c} -1, \cdots,$ and $N=N_\mathrm{c} +1$, add all these $N_\mathrm{c} $ equations together, and then divide the result by $N_\mathrm{c} $, we find that
\begin{eqnarray}
\label{eq:EeMinusT1}
 |E_\mathrm{e}-(T_1+T_{N_\mathrm{c}  }/N_\mathrm{c} -T_{2N_\mathrm{c}  }/N_\mathrm{c} )|<\epsilon/4.				\end{eqnarray}
If we apply Eq.~\ref{eq:Nb} for $N=N_\mathrm{c} $,
\begin{eqnarray}
\label{eq:EbMinusT1}
 |E_\mathrm{b}-(T_1-T_{N_\mathrm{c}  }/N_\mathrm{c} )|<\epsilon/4.					
\end{eqnarray}
Now, if we add Eq.~\ref{eq:EeMinusT1} and Eq.~\ref{eq:EbMinusT1},
\begin{eqnarray}
\label{eq:EePlusEb}
  |E_\mathrm{e}+E_\mathrm{b}-(2T_1-T_{2N_\mathrm{c}  }/N_\mathrm{c} )|<\epsilon/2.		
\end{eqnarray}
If we apply Eq.~\ref{eq:Nb} for $N=2N_\mathrm{c} $,
\begin{eqnarray}
\label{eq:twoEb}
|2E_\mathrm{b}-(2T_1-T_{2N_\mathrm{c}  }/N_\mathrm{c} )|<\epsilon/2.				\end{eqnarray}
If we subtract Eq.~\ref{eq:twoEb} from Eq.~\ref{eq:EePlusEb},
\begin{eqnarray}
\label{eq:EeMinusEb}
 |E_\mathrm{e}-E_\mathrm{b} |<\epsilon,
\end{eqnarray}
which completes the proof that $E_\mathrm{e}=E_\mathrm{b}$. We emphasize that in our general mathematical proof, we did not assume that the interactions are pairwise additive or decay fast in real space.

The proposed method of calculating the exfoliation energy has
several important advantages over the slab method.
First, our method is immensely efficient from
the computational point of view.
The energy of a layer in bulk
is obtained by a calculation on a primitive unit cell
and that of an exfoliated layer by a calculation on the supercell containing only
a single layer in vacuum.
Second, our method automatically and rigorously incorporates the effect of any surface reconstructions.
The slab method often ignores them to reduce the computational load.
Third, our method can easily take into account the relaxation of the in-plane lattice parameters 
of the exfoliated layer.
The slab method usually neglects this relaxation and uses
a supercell containing both the exfoliated layer and the remaining slab. 
Fourth, our method is compatible with any kind of many-body computational theories. 
We note that Hanke, \cite{Hanke2011}
Bj\"orkman {\it et}\,{\it al.}, \cite{Bjorkman2012}
and Chen {\it et}\,{\it al.}\ \cite{Chen2013}
have shown that, under the assumption that the interaction
among different layers is pairwise additive,
the exfoliation energy is the same as
the interlayer binding energy, i.\,e.\,, the energy (per layer) required to separate all the layers of the bulk.
However, from computational studies employing quantum Monte Carlo
simulations it is known that non-pairwise-additive many-body effects
such as Pauli repulsion and van der Waals interactions play an
important role in the interlayer interactions in layered
materials \cite{Spanu2009,Shulenburger2015,Mostaani2015,Dubecky2016}.
Our proposed method does not rely on such an assumption and may be combined with state-of-the-art
many-body techniques in accurately calculating
exfoliation energies.

Some studies (e.g., Ref.\ \cite{Bjorkman2012}) have justified the use of interlayer binding energy as an approximation to the exfoliation energy through an explicit convergence test of the exfoliation energy with respect to the number of atomic layers in the slab in limited cases.  Other studies (e.g., Ref.\ \cite{Sansone2016}) 
resort to a previous one \cite{Bjorkman2012} in using the equality of the exfoliation energy and layer binding energy.
The novelty of our study lies in proving the equality of the interlayer binding energy and the exfoliation energy independent of (i) the electronic structure methods, (ii) the energy functionals, and (iii) the class of materials being studied.  By doing so, our study enables people to focus only on improving the electronic structure methods and / or energy functionals for calculating the exfoliation energy of any material without worrying about the convergence with respect to the number of atomic layers in the slab.
So far, only the research groups with enough computational resources were able to do the slab calculations which were believed to be more accurate but computationally heavier than other methods and to check the convergence of their calculated exfoliation energies with respect to the number of atomic layers in the slab.  Research groups not accessible to high computational power had to resort to what has been believed to be an approximation. 
Our work can give a deep impact on this situation since we have proved that such an ``approximation'' of the exfoliation energy by the interlayer binding energy is not actually an approximation but is rigorous and exact.  Therefore, our study allows people not accessible to high-power supercomputing facilities, to have confidence in their calculations on surface-related energies.

Until now we have focused on simple layered materials
whose basis atoms can be contained in one atomic layer. 
In fact,
our method can be used to calculate the exfoliation energy of
any materials
whose surfaces before and after the exfoliation are equivalent,
i.\,e.\,, related by symmetry operations.
Most naturally occurring materials satisfy this criterion.
For example,
the remaining surface of the Bernal (AB)-stacked graphite after the
exfoliation of one graphene layer
is equivalent to the original surface after a 180$^\circ$ rotation with
respect to an axis normal to the surface.
In the case of rhombohedral graphite with the ABC stacking, \cite{Yang2001} 
the two surfaces before and after the exfoliation of a graphene layer
are connected by a simple translation.
Many layered materials meet this symmetry criterion.

Our method also applies to $n$-layer exfoliation (for $n>1$)
as long as the initial and final surfaces are equivalent.
The $n$-layer exfoliation energy per unit area $E_{\mathrm{exf}}(n)$ is given by
\begin{eqnarray}
E_{\mathrm{exf}}(n)=\frac{E_{\mathrm{iso}} (n)- E_{\mathrm{bulk}} n/m }{\mathcal{A}},
\label{eq:nlayer}
\end{eqnarray}
where $E_{\mathrm{iso}} (n)$ is the energy of the unit cell of an isolated $n$-layer slab in vacuum, $E_{\mathrm{bulk}}$ is the energy of the unit cell of a bulk material composed of $m$ layers (e.\,g.\,, $m = 2$ in the case of AB-stacked graphite or hBN), thus $E_{\mathrm{bulk}}/m$ corresponds to the energy of the bulk per layer, and $\mathcal{A}$ is the in-plane area of the bulk unit cell.

\begin{figure}[tbp]  
    \centering
    \includegraphics{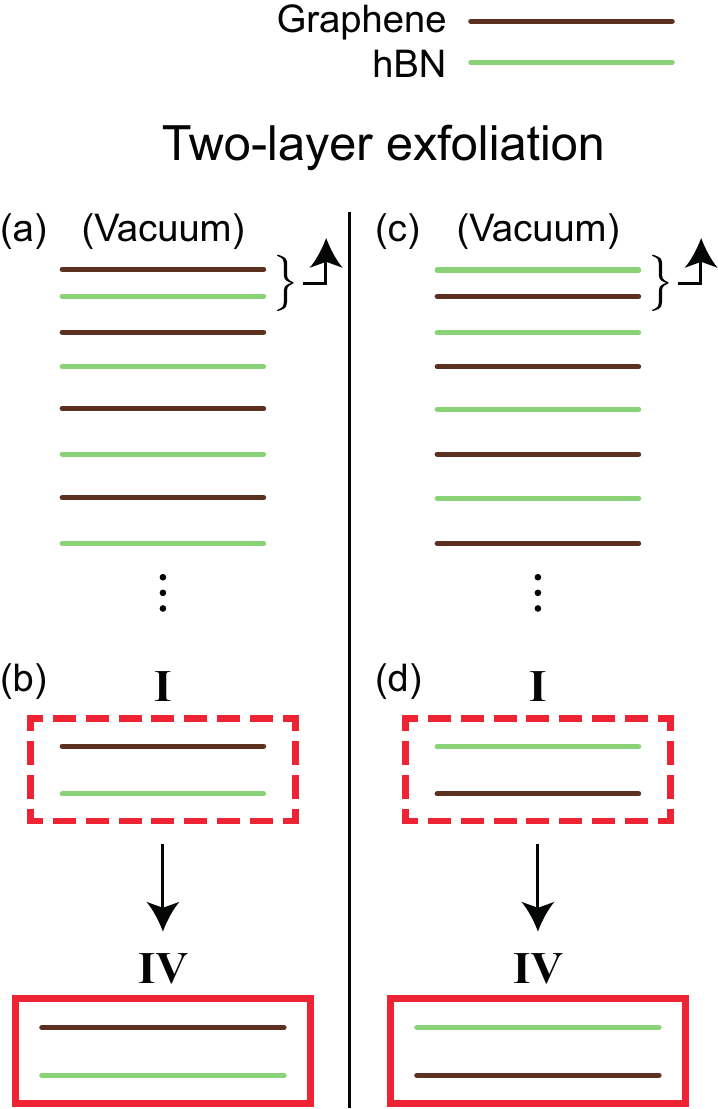}
\caption{\label{figure-graphene-BN}
Two different surfaces of graphene-hBN superlattice and the two-layer exfoliation processes therefrom.
(a) The exfoliation of top two layers from
(vacuum)-graphene-hBN-graphene-hBN-$\cdots$ surface.
(b) The net difference between configurations I and IV (presented in Fig.~\ref{figure-new}) in the two-layer exfoliation shown in (a).
(c) and (d) Similar quantities as in (a) and (b)
for (vacuum)-hBN-graphene-hBN-graphene-$\cdots$ surface.
    }
\end{figure}

Our finding also leads us to some interesting  results on the exfoliation energy. 
We consider a graphene-hBN superlattice,
which may have two types of surface terminations
(Figs.~\ref{figure-graphene-BN}(a) and~\ref{figure-graphene-BN}(c)). We predict that the two-layer exfoliation energies from
these surfaces are the same. The reason is that
the exfoliation energy is the difference in the energy of the two layers in bulk and that in vacuum, both of which in this case do not depend on the surface termination
(Figs.~\ref{figure-graphene-BN}(b) and~\ref{figure-graphene-BN}(d)).

For demonstration purposes,
we applied our proposed method of calculating
exfoliation energies from first principles
to representative 2D materials: graphene,  hBN,
 MoS$_2$,
and phosphorene.
The electronic structures were calculated using the
density-functional theory within the projector-augmented-wave method \cite{Kresse1999} as implemented in \textsc{vasp} \cite{Kresse1996}. 
The exchange-correlation energy was calculated using the functional of
Perdew, Burke, and Ernzerhof (PBE) \cite{Perdew1996}.
The material-dependent kinetic energy cutoffs were set to values higher
than 700~eV after convergence tests.
For {\bf k}-space integrations of bulk materials, a mesh of $19 \times 19 \times 7$ {\bf k} points was used for graphite, $19 \times 19 \times 7$ for hBN, $ 15 \times 15 \times 4$ for MoS$_2$, and $ 12 \times 9 \times 4$ for black phosphorus.
For {\it n}-layer slabs ($n=1\sim8$), we have
used the corresponding $M\times N\times1$ {\bf k}-point meshes.
We checked the convergence of the total energy with respect to
the number of {\bf k} points.
To avoid spurious interactions between periodic images,
a 13~\AA\ of vacuum was inserted between the {\it n}-layer slabs.
The geometry optimization was performed until an additional
ionic-relaxation step changed the total energy
per unit cell by less than $10^{-7}$~eV.
We took into account van der Waals interactions in the total energies by using the D2 scheme developed by Grimme \cite{Grimme2006}.
The dispersion corrections are sensitive to the specific functionals being used \cite{Bachhuber2015,Sansone2016}. 
However, to figure out which electronic-structure methods or energy functionals are good is beyond the purpose of our work, and we have not attempted to make comparison between different functionals.
We have checked that the results of our calculations are in agreement with those of other studies obtained from  the PBE-D2 exchange-correlation functional \footnote{The interlayer binding energies of graphite (21.2~meV/\AA$^2$) and black phosphorus (22.9~meV/\AA$^2$) obtained by neglecting the relaxation of the in-plane lattice parameters are in good agreement with those obtained in a similar way by Hanke et al.\ \cite{Hanke2011} for graphite, 21.8~meV/\AA$^2$, and by Sansone et al.\ \cite{Sansone2016} for black phosphorus, 23.0~meV/\AA$^2$.
}. 

\begin{figure}[tbp]  
    \centering
    \includegraphics{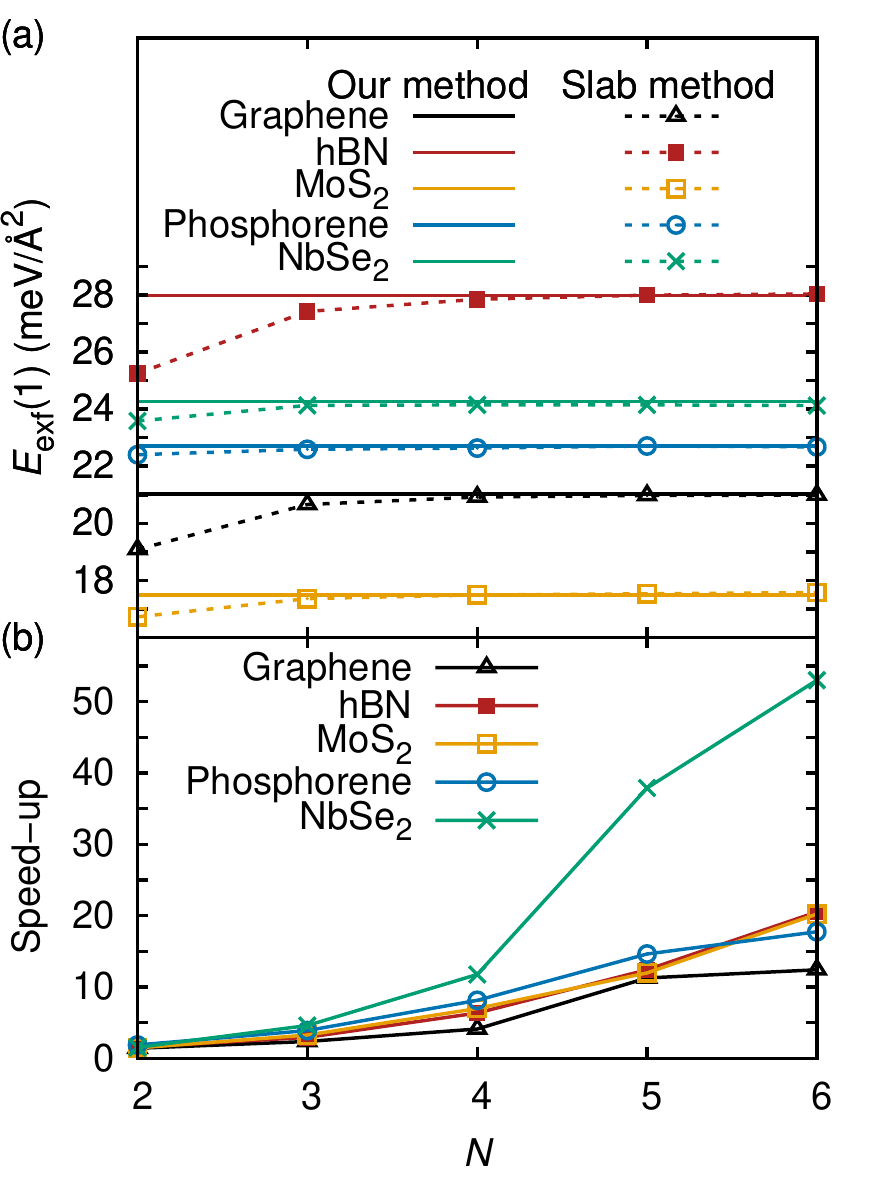}
    \caption{\label{figure-efficiency} 
Comparison of the slab method and our proposed method.  (a) Monolayer exfoliation energies obtained by the $N$-layer slab method and those obtained by the proposed method.  (b) The ratio of the required time for computing the monolayer exfoliation energy by the $N$-layer slab method to that by our method.
    }
\end{figure}

We now compare the computation time of our method with that of the slab method as presented in Fig.~\ref{figure-efficiency}. For this purpose, we assumed that the exfoliated atomic layer has the same in-plane lattice constants as the bulk. 
We performed calculations on all the materials that we discussed before and NbSe$_2$, a metallic layer-structured material. 
Since NbSe$_2$ is a metal, we  used dense {\bf k}-point meshes of $21 \times 21 \times 6$ and $21 \times 21 \times 1$ for the bulk material and for the slabs, respectively, and neglected the possible reconstructions of the bulk, surface, and monolayer of NbSe$_2$ which are still controversial \cite{Malliakas2013,Langer2014,Ugeda2016,Silva2016}. For other materials, we used the same {\bf k}-meshes
as adopted in calculations using our method.

As we expected, the calculated exfoliation energies obtained from the slab method converge with $N$ to those obtained from our proposed method (Fig.~\ref{figure-efficiency}(a)).  Although it depends on the kind of the material, we need $N=4$ for a convergence of the monolayer exfoliation energy within a few percent.  The required computation time for the exfoliation energy of our method is a few times or even ten times shorter than that of the slab method (Fig.~\ref{figure-efficiency}(b)).  It is to be noted that both (i) the minimum number of atomic layers in the slab and (ii) the computation time for converging the slab-method calculations do depend on the electronic-structure method, the energy functional, and the material.  Even if we use the same density-functional-theory scheme and the same exchange-correlation functional, the advantage of our method for not using a thick slab of atomic layers will be greater if the material itself has bulk or surface reconstructions. Moreover, if our method is combined with more advanced many-body techniques, the advantages of our method over the slab method would be even greater since the scaling of the computation time with respect to the total number of electrons in those many-body methods is much worse. For example, our method can be directly combined with quantum-chemistry techniques for periodic crystals using many-body wavefunctions \cite{Booth2013}.

Figure~\ref{figure-efficiency} also shows that the energy convergence (as a function of the number of layers $N$) of the slab method for this metallic material is similar to that of semiconducting systems.  On the other hand, the ratio of the computational time required for the slab method to our method is much higher in the case of a metal than a semiconductor, i.e., the advantage of our approach over the slab method is much greater for a metallic system. 
The convergence of calculations with respect to the total number of layers in a slab may be slow, and especially so if long-range dispersion interactions are taken into account.  Our new approach is exempt from this convergence test because we need to calculate only two configurations, the bulk and the isolated single layer. (As evident from our mathematical proof, the computational complications originated from the existence of the surface in reality exactly cancel out between two geometries, namely, before and after the exfoliation.)  Therefore, although the calculation would necessarily become heavy with the correlated wave-function methods in all cases including our new approach, the reduction in the computational load with our new method is tremendous compared to the conventional slab methods.

\begin{figure}[tbp]  
    \centering
    \includegraphics{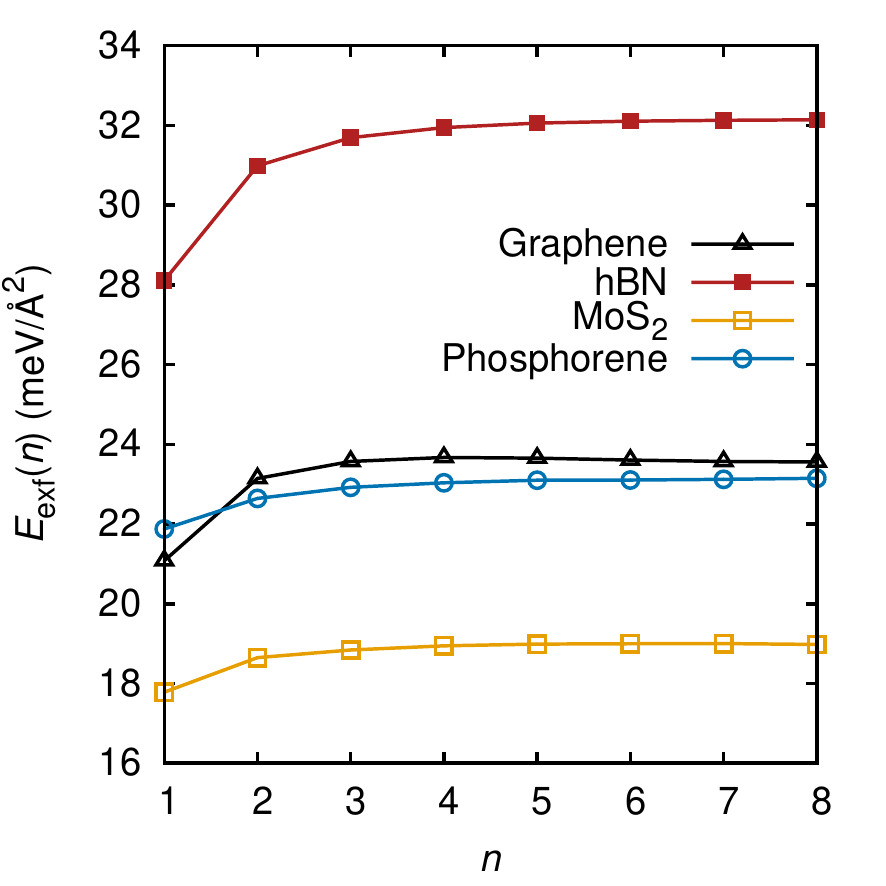}
    \caption{\label{figure-exfoliation} 
    $n$-layer exfoliation energies per area, $E_{\mathrm{exf}}(n)$ (see Eq.~(\ref{eq:nlayer})), of graphene, hBN, MoS$_2$ and phosphorene.
    }
\end{figure}

Figure~\ref{figure-exfoliation} shows that the
$n$-layer exfoliation energy,
$E_{\mathrm{exf}} (n)$ [Eq.~\eqref{eq:nlayer}],
of the four materials
calculated from our proposed method
is in the range of 18--32 meV/\AA$^2$. 
$E_{\mathrm{exf}} (n)$
increases with $n$ since
the surface $n$-layer slab with higher $n$ contains more atoms
and, consequently, its total interaction with the bulk underneath before exfoliation
is stronger.
In the following analysis, we regard $E_{\mathrm{exf}} (8)$
as the limiting value $E_{\mathrm{exf}} (\infty)$, which seems a good approximation from the convergence trend in Fig.~\ref{figure-exfoliation}.
Note that $E_{\mathrm{exf}} (\infty)$ is, by definition, the {\it cleavage energy},
the energy required to split a bulk crystal into two bulk
parts \cite{Whittaker1982}. 
In the case of graphene, MoS$_2$, and phosphorene, 
$E_{\mathrm{exf}} (n)$ converges within 1~\%
of $E_{\mathrm{exf}} (\infty)$ when $n\geq3$. For hBN, the convergence of $E_{\mathrm{exf}} (n)$
within 1~\% is reached when $n\geq4$.
We also note that
the difference of the cleavage energy of graphene from its one-layer exfoliation energy [$E_{\mathrm{exf}}(1)$]
is 2.5~meV/\AA$^2$, which is 12~\% of the one-layer exfoliation energy.
The corresponding quantities are 4.0~meV/\AA$^2$ and 14~\% for hBN,
1.2~meV/\AA$^2$ and 6.7~\% for MoS$_2$, and
1.3~meV/\AA$^2$ and 5.8~\% for phosphorene.

\begin{figure*}[tbp]  
    \centering
    \includegraphics{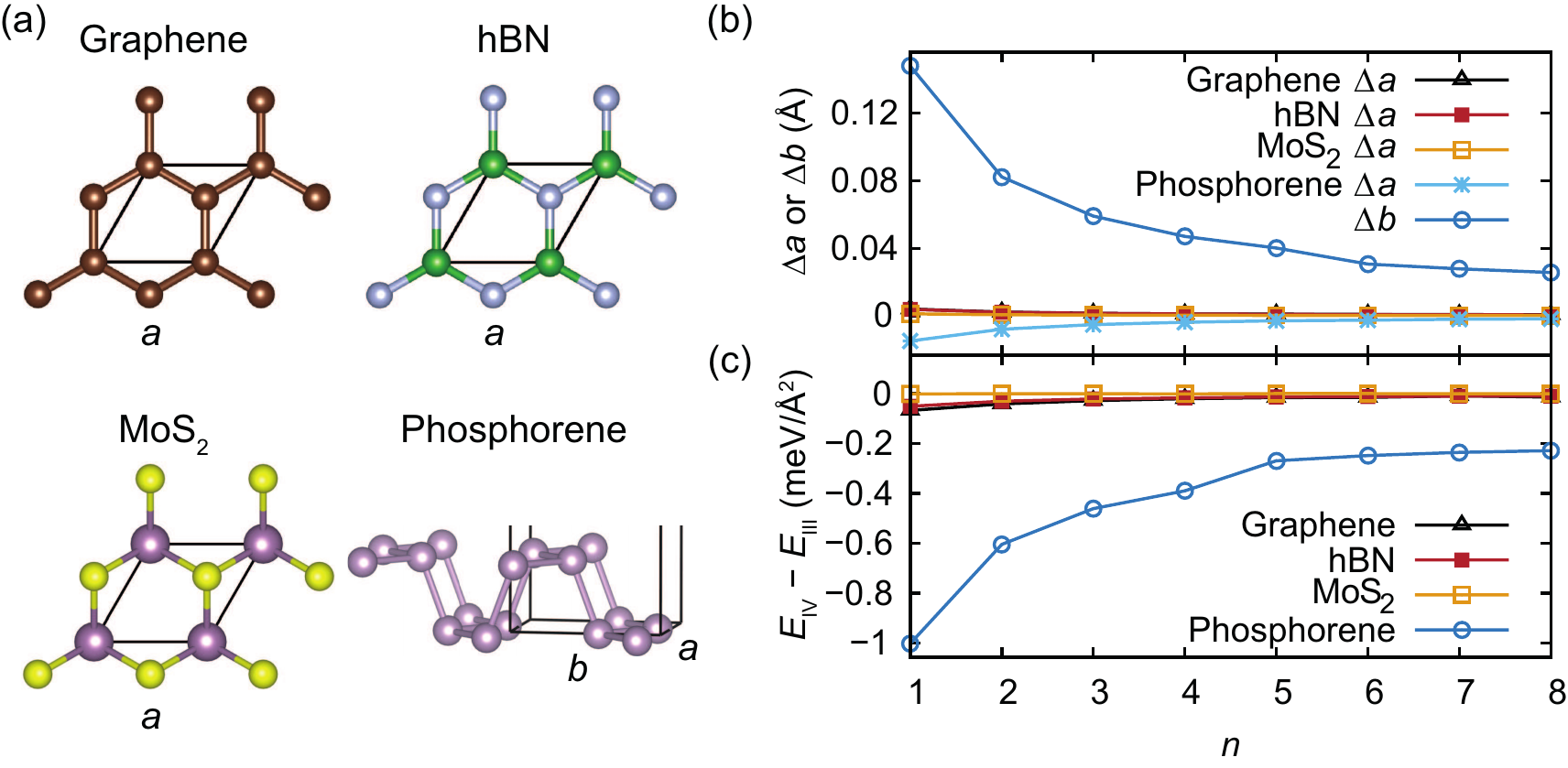}
    \caption{\label{figure-in-plane}
    (a) The in-plane lattice parameters of the materials considered.
    (b) The difference in the in-plane lattice parameter(s) of an $n$-layer slab  and a bulk
crystal.  (c) The stabilization energy ($E_{\mathrm{IV}}-E_{\mathrm{III}}$) of the exfoliated layer(s) by the relaxation of in-plane lattice constants.
    }
\end{figure*}

So far, we have demonstrated how our method can be used to calculate
the exfoliation energies of {\it n}-atomic-layer slabs.
Now we focus on a particular step of the exfoliation process,
the lattice constant change and the subsequent structural relaxation of the exfoliated $n$-layer slab (III$\to$IV in Fig.~\ref{figure-new}), 
which was usually not treated accurately 
in conventional methods since they employ calculations on a supercell containing both the exfoliated layer and the remaining slab. 
As the number of layers increases, the change in the lattice constants from those of the bulk
diminishes (Fig.~\ref{figure-in-plane}).
Among the four materials, the $b$ parameter of phosphorene (Fig.~\ref{figure-in-plane}(a)) shows the largest change (increase) in the lattice constants upon relaxation: $\Delta b / b = 3.3$~\% (Fig.~\ref{figure-in-plane}(b)).
This large change is due to the structural compliance of phosphorene in the armchair
direction.
The change in the lattice constants
stabilizes the exfoliated {\it n}-layer slab as shown in
Figs.~\ref{figure-in-plane}(b) and~\ref{figure-in-plane}(c).
Among the four materials studied, the stabilization energy ($E_{\mathrm{IV}}-E_{\mathrm{III}}$) of phosphorene is the largest.
Figure~\ref{figure-in-plane}(c) shows that $E_{\mathrm{IV}}-E_{\mathrm{III}}$ (defined to be negative here) accounts for 5~\% of the exfoliation energy
of monolayer phosphorene, whereas this contribution is
less than 0.4~\% for the other materials.

In conclusion, we developed an extremely efficient method
to calculate the exfoliation energy from first 
principles. This scheme is also capable of 
taking into account surface relaxations and reconstructions {\it rigorously}, handling the change in the in-plane lattice constants of exfoliated layers easily, and being easily combined with state-of-the-art many-body techniques.
We applied this method to calculate {\it n}-layer exfoliation
energies of four representative 2D materials.
Although our demonstration is focused on layered materials with van der Waals interactions,
the method can in general be applied to other classes of layered materials, e.\,g.\,,
those with hydrogen bonds.
We believe that because of the unique merits of our method it will replace the traditional method and be widely
used to calculate mechanical exfoliation energies of 2D materials.

\begin{acknowledgments}
This work was supported by the Creative-Pioneering Research Program through Seoul National University.
Computational resources were provided by the Korea Institute of Science and Technology Information.
\end{acknowledgments}

\bibliography{reference}

\providecommand{\noopsort}[1]{}\providecommand{\singleletter}[1]{#1}%
\begin{thebibliography}{41}%
\makeatletter
\providecommand \@ifxundefined [1]{%
 \@ifx{#1\undefined}
}%
\providecommand \@ifnum [1]{%
 \ifnum #1\expandafter \@firstoftwo
 \else \expandafter \@secondoftwo
 \fi
}%
\providecommand \@ifx [1]{%
 \ifx #1\expandafter \@firstoftwo
 \else \expandafter \@secondoftwo
 \fi
}%
\providecommand \natexlab [1]{#1}%
\providecommand \enquote  [1]{``#1''}%
\providecommand \bibnamefont  [1]{#1}%
\providecommand \bibfnamefont [1]{#1}%
\providecommand \citenamefont [1]{#1}%
\providecommand \href@noop [0]{\@secondoftwo}%
\providecommand \href [0]{\begingroup \@sanitize@url \@href}%
\providecommand \@href[1]{\@@startlink{#1}\@@href}%
\providecommand \@@href[1]{\endgroup#1\@@endlink}%
\providecommand \@sanitize@url [0]{\catcode `\\12\catcode `\$12\catcode
  `\&12\catcode `\#12\catcode `\^12\catcode `\_12\catcode `\%12\relax}%
\providecommand \@@startlink[1]{}%
\providecommand \@@endlink[0]{}%
\providecommand \url  [0]{\begingroup\@sanitize@url \@url }%
\providecommand \@url [1]{\endgroup\@href {#1}{\urlprefix }}%
\providecommand \urlprefix  [0]{URL }%
\providecommand \Eprint [0]{\href }%
\providecommand \doibase [0]{http://dx.doi.org/}%
\providecommand \selectlanguage [0]{\@gobble}%
\providecommand \bibinfo  [0]{\@secondoftwo}%
\providecommand \bibfield  [0]{\@secondoftwo}%
\providecommand \translation [1]{[#1]}%
\providecommand \BibitemOpen [0]{}%
\providecommand \bibitemStop [0]{}%
\providecommand \bibitemNoStop [0]{.\EOS\space}%
\providecommand \EOS [0]{\spacefactor3000\relax}%
\providecommand \BibitemShut  [1]{\csname bibitem#1\endcsname}%
\let\auto@bib@innerbib\@empty
\bibitem [{\citenamefont {Novoselov}\ \emph {et~al.}(2004)\citenamefont
  {Novoselov}, \citenamefont {Geim}, \citenamefont {Morozov}, \citenamefont
  {Jiang}, \citenamefont {Zhang}, \citenamefont {Dubonos}, \citenamefont
  {Grigorieva},\ and\ \citenamefont {Firsov}}]{Novoselov2004}%
  \BibitemOpen
  \bibfield  {author} {\bibinfo {author} {\bibfnamefont {K.~S.}\ \bibnamefont
  {Novoselov}}, \bibinfo {author} {\bibfnamefont {A.~K.}\ \bibnamefont {Geim}},
  \bibinfo {author} {\bibfnamefont {S.~V.}\ \bibnamefont {Morozov}}, \bibinfo
  {author} {\bibfnamefont {D.}~\bibnamefont {Jiang}}, \bibinfo {author}
  {\bibfnamefont {Y.}~\bibnamefont {Zhang}}, \bibinfo {author} {\bibfnamefont
  {S.~V.}\ \bibnamefont {Dubonos}}, \bibinfo {author} {\bibfnamefont {I.~V.}\
  \bibnamefont {Grigorieva}}, \ and\ \bibinfo {author} {\bibfnamefont {A.~A.}\
  \bibnamefont {Firsov}},\ }\href@noop {} {\bibfield  {journal} {\bibinfo
  {journal} {Science}\ }\textbf {\bibinfo {volume} {306}},\ \bibinfo {pages}
  {666} (\bibinfo {year} {2004})}\BibitemShut {NoStop}%
\bibitem [{\citenamefont {Novoselov}\ \emph {et~al.}(2016)\citenamefont
  {Novoselov}, \citenamefont {Mishchenko}, \citenamefont {Carvalho},\ and\
  \citenamefont {Neto}}]{Novoselov2016}%
  \BibitemOpen
  \bibfield  {author} {\bibinfo {author} {\bibfnamefont {K.}~\bibnamefont
  {Novoselov}}, \bibinfo {author} {\bibfnamefont {A.}~\bibnamefont
  {Mishchenko}}, \bibinfo {author} {\bibfnamefont {A.}~\bibnamefont
  {Carvalho}}, \ and\ \bibinfo {author} {\bibfnamefont {A.~C.}\ \bibnamefont
  {Neto}},\ }\href@noop {} {\bibfield  {journal} {\bibinfo  {journal}
  {Science}\ }\textbf {\bibinfo {volume} {353}},\ \bibinfo {pages} {aac9439}
  (\bibinfo {year} {2016})}\BibitemShut {NoStop}%
\bibitem [{\citenamefont {Ferrari}\ \emph {et~al.}(2015)\citenamefont
  {Ferrari}, \citenamefont {Bonaccorso}, \citenamefont {Fal'ko}, \citenamefont
  {Novoselov}, \citenamefont {Roche}, \citenamefont {Boggild}, \citenamefont
  {Borini}, \citenamefont {Koppens}, \citenamefont {Palermo}, \citenamefont
  {Pugno}, \citenamefont {Garrido}, \citenamefont {Sordan}, \citenamefont
  {Bianco}, \citenamefont {Ballerini}, \citenamefont {Prato}, \citenamefont
  {Lidorikis}, \citenamefont {Kivioja}, \citenamefont {Marinelli},
  \citenamefont {Ryhaenen}, \citenamefont {Morpurgo}, \citenamefont {Coleman},
  \citenamefont {Nicolosi}, \citenamefont {Colombo}, \citenamefont {Fert},
  \citenamefont {Garcia-Hernandez}, \citenamefont {Bachtold}, \citenamefont
  {Schneider}, \citenamefont {Guinea}, \citenamefont {Dekker}, \citenamefont
  {Barbone}, \citenamefont {Sun}, \citenamefont {Galiotis}, \citenamefont
  {Grigorenko}, \citenamefont {Konstantatos}, \citenamefont {Kis},
  \citenamefont {Katsnelson}, \citenamefont {Vandersypen}, \citenamefont
  {Loiseau}, \citenamefont {Morandi}, \citenamefont {Neumaier}, \citenamefont
  {Treossi}, \citenamefont {Pellegrini}, \citenamefont {Polini}, \citenamefont
  {Tredicucci}, \citenamefont {Williams}, \citenamefont {Hong}, \citenamefont
  {Ahn}, \citenamefont {Kim}, \citenamefont {Zirath}, \citenamefont {van Wees},
  \citenamefont {van~der Zant}, \citenamefont {Occhipinti}, \citenamefont
  {Di~Matteo}, \citenamefont {Kinloch}, \citenamefont {Seyller}, \citenamefont
  {Quesnel}, \citenamefont {Feng}, \citenamefont {Teo}, \citenamefont
  {Rupesinghe}, \citenamefont {Hakonen}, \citenamefont {Neil}, \citenamefont
  {Tannock}, \citenamefont {Loefwander},\ and\ \citenamefont
  {Kinaret}}]{Ferrari2015}%
  \BibitemOpen
  \bibfield  {author} {\bibinfo {author} {\bibfnamefont {A.~C.}\ \bibnamefont
  {Ferrari}}, \bibinfo {author} {\bibfnamefont {F.}~\bibnamefont {Bonaccorso}},
  \bibinfo {author} {\bibfnamefont {V.}~\bibnamefont {Fal'ko}}, \bibinfo
  {author} {\bibfnamefont {K.~S.}\ \bibnamefont {Novoselov}}, \bibinfo {author}
  {\bibfnamefont {S.}~\bibnamefont {Roche}}, \bibinfo {author} {\bibfnamefont
  {P.}~\bibnamefont {Boggild}}, \bibinfo {author} {\bibfnamefont
  {S.}~\bibnamefont {Borini}}, \bibinfo {author} {\bibfnamefont {F.~H.~L.}\
  \bibnamefont {Koppens}}, \bibinfo {author} {\bibfnamefont {V.}~\bibnamefont
  {Palermo}}, \bibinfo {author} {\bibfnamefont {N.}~\bibnamefont {Pugno}},
  \bibinfo {author} {\bibfnamefont {J.~A.}\ \bibnamefont {Garrido}}, \bibinfo
  {author} {\bibfnamefont {R.}~\bibnamefont {Sordan}}, \bibinfo {author}
  {\bibfnamefont {A.}~\bibnamefont {Bianco}}, \bibinfo {author} {\bibfnamefont
  {L.}~\bibnamefont {Ballerini}}, \bibinfo {author} {\bibfnamefont
  {M.}~\bibnamefont {Prato}}, \bibinfo {author} {\bibfnamefont
  {E.}~\bibnamefont {Lidorikis}}, \bibinfo {author} {\bibfnamefont
  {J.}~\bibnamefont {Kivioja}}, \bibinfo {author} {\bibfnamefont
  {C.}~\bibnamefont {Marinelli}}, \bibinfo {author} {\bibfnamefont
  {T.}~\bibnamefont {Ryhaenen}}, \bibinfo {author} {\bibfnamefont
  {A.}~\bibnamefont {Morpurgo}}, \bibinfo {author} {\bibfnamefont {J.~N.}\
  \bibnamefont {Coleman}}, \bibinfo {author} {\bibfnamefont {V.}~\bibnamefont
  {Nicolosi}}, \bibinfo {author} {\bibfnamefont {L.}~\bibnamefont {Colombo}},
  \bibinfo {author} {\bibfnamefont {A.}~\bibnamefont {Fert}}, \bibinfo {author}
  {\bibfnamefont {M.}~\bibnamefont {Garcia-Hernandez}}, \bibinfo {author}
  {\bibfnamefont {A.}~\bibnamefont {Bachtold}}, \bibinfo {author}
  {\bibfnamefont {G.~F.}\ \bibnamefont {Schneider}}, \bibinfo {author}
  {\bibfnamefont {F.}~\bibnamefont {Guinea}}, \bibinfo {author} {\bibfnamefont
  {C.}~\bibnamefont {Dekker}}, \bibinfo {author} {\bibfnamefont
  {M.}~\bibnamefont {Barbone}}, \bibinfo {author} {\bibfnamefont
  {Z.}~\bibnamefont {Sun}}, \bibinfo {author} {\bibfnamefont {C.}~\bibnamefont
  {Galiotis}}, \bibinfo {author} {\bibfnamefont {A.~N.}\ \bibnamefont
  {Grigorenko}}, \bibinfo {author} {\bibfnamefont {G.}~\bibnamefont
  {Konstantatos}}, \bibinfo {author} {\bibfnamefont {A.}~\bibnamefont {Kis}},
  \bibinfo {author} {\bibfnamefont {M.}~\bibnamefont {Katsnelson}}, \bibinfo
  {author} {\bibfnamefont {L.}~\bibnamefont {Vandersypen}}, \bibinfo {author}
  {\bibfnamefont {A.}~\bibnamefont {Loiseau}}, \bibinfo {author} {\bibfnamefont
  {V.}~\bibnamefont {Morandi}}, \bibinfo {author} {\bibfnamefont
  {D.}~\bibnamefont {Neumaier}}, \bibinfo {author} {\bibfnamefont
  {E.}~\bibnamefont {Treossi}}, \bibinfo {author} {\bibfnamefont
  {V.}~\bibnamefont {Pellegrini}}, \bibinfo {author} {\bibfnamefont
  {M.}~\bibnamefont {Polini}}, \bibinfo {author} {\bibfnamefont
  {A.}~\bibnamefont {Tredicucci}}, \bibinfo {author} {\bibfnamefont {G.~M.}\
  \bibnamefont {Williams}}, \bibinfo {author} {\bibfnamefont {B.~H.}\
  \bibnamefont {Hong}}, \bibinfo {author} {\bibfnamefont {J.-H.}\ \bibnamefont
  {Ahn}}, \bibinfo {author} {\bibfnamefont {J.~M.}\ \bibnamefont {Kim}},
  \bibinfo {author} {\bibfnamefont {H.}~\bibnamefont {Zirath}}, \bibinfo
  {author} {\bibfnamefont {B.~J.}\ \bibnamefont {van Wees}}, \bibinfo {author}
  {\bibfnamefont {H.}~\bibnamefont {van~der Zant}}, \bibinfo {author}
  {\bibfnamefont {L.}~\bibnamefont {Occhipinti}}, \bibinfo {author}
  {\bibfnamefont {A.}~\bibnamefont {Di~Matteo}}, \bibinfo {author}
  {\bibfnamefont {I.~A.}\ \bibnamefont {Kinloch}}, \bibinfo {author}
  {\bibfnamefont {T.}~\bibnamefont {Seyller}}, \bibinfo {author} {\bibfnamefont
  {E.}~\bibnamefont {Quesnel}}, \bibinfo {author} {\bibfnamefont
  {X.}~\bibnamefont {Feng}}, \bibinfo {author} {\bibfnamefont {K.}~\bibnamefont
  {Teo}}, \bibinfo {author} {\bibfnamefont {N.}~\bibnamefont {Rupesinghe}},
  \bibinfo {author} {\bibfnamefont {P.}~\bibnamefont {Hakonen}}, \bibinfo
  {author} {\bibfnamefont {S.~R.~T.}\ \bibnamefont {Neil}}, \bibinfo {author}
  {\bibfnamefont {Q.}~\bibnamefont {Tannock}}, \bibinfo {author} {\bibfnamefont
  {T.}~\bibnamefont {Loefwander}}, \ and\ \bibinfo {author} {\bibfnamefont
  {J.}~\bibnamefont {Kinaret}},\ }\href {\doibase {10.1039/c4nr01600a}}
  {\bibfield  {journal} {\bibinfo  {journal} {{Nanoscale}}\ }\textbf {\bibinfo
  {volume} {{7}}},\ \bibinfo {pages} {4598} (\bibinfo {year}
  {{2015}})}\BibitemShut {NoStop}%
\bibitem [{\citenamefont {Mounet}\ \emph {et~al.}(2018)\citenamefont {Mounet},
  \citenamefont {Gibertini}, \citenamefont {Schwaller}, \citenamefont {Campi},
  \citenamefont {Merkys}, \citenamefont {Marrazzo}, \citenamefont {Sohier},
  \citenamefont {Castelli}, \citenamefont {Cepellotti}, \citenamefont {Pizzi},\
  and\ \citenamefont {Marzari}}]{Mounet2018}%
  \BibitemOpen
  \bibfield  {author} {\bibinfo {author} {\bibfnamefont {N.}~\bibnamefont
  {Mounet}}, \bibinfo {author} {\bibfnamefont {M.}~\bibnamefont {Gibertini}},
  \bibinfo {author} {\bibfnamefont {P.}~\bibnamefont {Schwaller}}, \bibinfo
  {author} {\bibfnamefont {D.}~\bibnamefont {Campi}}, \bibinfo {author}
  {\bibfnamefont {A.}~\bibnamefont {Merkys}}, \bibinfo {author} {\bibfnamefont
  {A.}~\bibnamefont {Marrazzo}}, \bibinfo {author} {\bibfnamefont
  {T.}~\bibnamefont {Sohier}}, \bibinfo {author} {\bibfnamefont {I.~E.}\
  \bibnamefont {Castelli}}, \bibinfo {author} {\bibfnamefont {A.}~\bibnamefont
  {Cepellotti}}, \bibinfo {author} {\bibfnamefont {G.}~\bibnamefont {Pizzi}}, \
  and\ \bibinfo {author} {\bibfnamefont {N.}~\bibnamefont {Marzari}},\ }\href
  {\doibase 10.1038/s41565-017-0035-5} {\bibfield  {journal} {\bibinfo
  {journal} {Nat. Nanotechnol.}\ }\textbf {\bibinfo {volume} {13}},\ \bibinfo
  {pages} {246} (\bibinfo {year} {2018})}\BibitemShut {NoStop}%
\bibitem [{\citenamefont {Peng}\ \emph {et~al.}(2016)\citenamefont {Peng},
  \citenamefont {Yang}, \citenamefont {Perdew},\ and\ \citenamefont
  {Sun}}]{Peng2016}%
  \BibitemOpen
  \bibfield  {author} {\bibinfo {author} {\bibfnamefont {H.}~\bibnamefont
  {Peng}}, \bibinfo {author} {\bibfnamefont {Z.-H.}\ \bibnamefont {Yang}},
  \bibinfo {author} {\bibfnamefont {J.~P.}\ \bibnamefont {Perdew}}, \ and\
  \bibinfo {author} {\bibfnamefont {J.}~\bibnamefont {Sun}},\ }\href {\doibase
  10.1103/PhysRevX.6.041005} {\bibfield  {journal} {\bibinfo  {journal} {Phys.
  Rev. X}\ }\textbf {\bibinfo {volume} {6}},\ \bibinfo {pages} {041005}
  (\bibinfo {year} {2016})}\BibitemShut {NoStop}%
\bibitem [{\citenamefont {Leb\`egue}\ \emph {et~al.}(2010)\citenamefont
  {Leb\`egue}, \citenamefont {Harl}, \citenamefont {Gould}, \citenamefont
  {\'Angy\'an}, \citenamefont {Kresse},\ and\ \citenamefont
  {Dobson}}]{Lebegue2010}%
  \BibitemOpen
  \bibfield  {author} {\bibinfo {author} {\bibfnamefont {S.}~\bibnamefont
  {Leb\`egue}}, \bibinfo {author} {\bibfnamefont {J.}~\bibnamefont {Harl}},
  \bibinfo {author} {\bibfnamefont {T.}~\bibnamefont {Gould}}, \bibinfo
  {author} {\bibfnamefont {J.~G.}\ \bibnamefont {\'Angy\'an}}, \bibinfo
  {author} {\bibfnamefont {G.}~\bibnamefont {Kresse}}, \ and\ \bibinfo {author}
  {\bibfnamefont {J.~F.}\ \bibnamefont {Dobson}},\ }\href {\doibase
  10.1103/PhysRevLett.105.196401} {\bibfield  {journal} {\bibinfo  {journal}
  {Phys. Rev. Lett.}\ }\textbf {\bibinfo {volume} {105}},\ \bibinfo {pages}
  {196401} (\bibinfo {year} {2010})}\BibitemShut {NoStop}%
\bibitem [{\citenamefont {Wang}\ \emph {et~al.}(2015)\citenamefont {Wang},
  \citenamefont {Dai}, \citenamefont {Li}, \citenamefont {Yang}, \citenamefont
  {Srolovitz},\ and\ \citenamefont {Zheng}}]{Wang2015}%
  \BibitemOpen
  \bibfield  {author} {\bibinfo {author} {\bibfnamefont {W.}~\bibnamefont
  {Wang}}, \bibinfo {author} {\bibfnamefont {S.}~\bibnamefont {Dai}}, \bibinfo
  {author} {\bibfnamefont {X.}~\bibnamefont {Li}}, \bibinfo {author}
  {\bibfnamefont {J.}~\bibnamefont {Yang}}, \bibinfo {author} {\bibfnamefont
  {D.~J.}\ \bibnamefont {Srolovitz}}, \ and\ \bibinfo {author} {\bibfnamefont
  {Q.}~\bibnamefont {Zheng}},\ }\href {\doibase 10.1038/ncomms8853} {\bibfield
  {journal} {\bibinfo  {journal} {Nat. Commun.}\ }\textbf {\bibinfo {volume}
  {6}},\ \bibinfo {pages} {7853} (\bibinfo {year} {2015})}\BibitemShut
  {NoStop}%
\bibitem [{\citenamefont {Gould}\ \emph {et~al.}(2013)\citenamefont {Gould},
  \citenamefont {Leb\`egue},\ and\ \citenamefont {Dobson}}]{Gould2013}%
  \BibitemOpen
  \bibfield  {author} {\bibinfo {author} {\bibfnamefont {T.}~\bibnamefont
  {Gould}}, \bibinfo {author} {\bibfnamefont {S.}~\bibnamefont {Leb\`egue}}, \
  and\ \bibinfo {author} {\bibfnamefont {J.~F.}\ \bibnamefont {Dobson}},\
  }\href {\doibase 10.1088/0953-8984/25/44/445010} {\bibfield  {journal}
  {\bibinfo  {journal} {J. Phys.: Condens. Matter}\ }\textbf {\bibinfo {volume}
  {25}},\ \bibinfo {pages} {445010} (\bibinfo {year} {2013})}\BibitemShut
  {NoStop}%
\bibitem [{\citenamefont {Liu}\ \emph {et~al.}(2012)\citenamefont {Liu},
  \citenamefont {Liu}, \citenamefont {Cheng}, \citenamefont {Li}, \citenamefont
  {Wang},\ and\ \citenamefont {Zheng}}]{Liu2012}%
  \BibitemOpen
  \bibfield  {author} {\bibinfo {author} {\bibfnamefont {Z.}~\bibnamefont
  {Liu}}, \bibinfo {author} {\bibfnamefont {J.~Z.}\ \bibnamefont {Liu}},
  \bibinfo {author} {\bibfnamefont {Y.}~\bibnamefont {Cheng}}, \bibinfo
  {author} {\bibfnamefont {Z.}~\bibnamefont {Li}}, \bibinfo {author}
  {\bibfnamefont {L.}~\bibnamefont {Wang}}, \ and\ \bibinfo {author}
  {\bibfnamefont {Q.}~\bibnamefont {Zheng}},\ }\href {\doibase
  10.1103/PhysRevB.85.205418} {\bibfield  {journal} {\bibinfo  {journal} {Phys.
  Rev. B}\ }\textbf {\bibinfo {volume} {85}},\ \bibinfo {pages} {205418}
  (\bibinfo {year} {2012})}\BibitemShut {NoStop}%
\bibitem [{\citenamefont {Spanu}\ \emph {et~al.}(2009)\citenamefont {Spanu},
  \citenamefont {Sorella},\ and\ \citenamefont {Galli}}]{Spanu2009}%
  \BibitemOpen
  \bibfield  {author} {\bibinfo {author} {\bibfnamefont {L.}~\bibnamefont
  {Spanu}}, \bibinfo {author} {\bibfnamefont {S.}~\bibnamefont {Sorella}}, \
  and\ \bibinfo {author} {\bibfnamefont {G.}~\bibnamefont {Galli}},\ }\href
  {\doibase 10.1103/PhysRevLett.103.196401} {\bibfield  {journal} {\bibinfo
  {journal} {Phys. Rev. Lett.}\ }\textbf {\bibinfo {volume} {103}},\ \bibinfo
  {pages} {1} (\bibinfo {year} {2009})}\BibitemShut {NoStop}%
\bibitem [{\citenamefont {Ziambaras}\ \emph {et~al.}(2007)\citenamefont
  {Ziambaras}, \citenamefont {Kleis}, \citenamefont {Schr\"oder},\ and\
  \citenamefont {Hyldgaard}}]{Ziambaras2007}%
  \BibitemOpen
  \bibfield  {author} {\bibinfo {author} {\bibfnamefont {E.}~\bibnamefont
  {Ziambaras}}, \bibinfo {author} {\bibfnamefont {J.}~\bibnamefont {Kleis}},
  \bibinfo {author} {\bibfnamefont {E.}~\bibnamefont {Schr\"oder}}, \ and\
  \bibinfo {author} {\bibfnamefont {P.}~\bibnamefont {Hyldgaard}},\ }\href
  {\doibase 10.1103/PhysRevB.76.155425} {\bibfield  {journal} {\bibinfo
  {journal} {Phys. Rev. B}\ }\textbf {\bibinfo {volume} {76}},\ \bibinfo
  {pages} {155425} (\bibinfo {year} {2007})}\BibitemShut {NoStop}%
\bibitem [{\citenamefont {Hanke}(2011)}]{Hanke2011}%
  \BibitemOpen
  \bibfield  {author} {\bibinfo {author} {\bibfnamefont {F.}~\bibnamefont
  {Hanke}},\ }\href {\doibase 10.1002/jcc.21724} {\bibfield  {journal}
  {\bibinfo  {journal} {J. Comput. Chem.}\ }\textbf {\bibinfo {volume} {32}},\
  \bibinfo {pages} {1424} (\bibinfo {year} {2011})}\BibitemShut {NoStop}%
\bibitem [{\citenamefont {Bj\"orkman}\ \emph {et~al.}(2012)\citenamefont
  {Bj\"orkman}, \citenamefont {Gulans}, \citenamefont {Krasheninnikov},\ and\
  \citenamefont {Nieminen}}]{Bjorkman2012}%
  \BibitemOpen
  \bibfield  {author} {\bibinfo {author} {\bibfnamefont {T.}~\bibnamefont
  {Bj\"orkman}}, \bibinfo {author} {\bibfnamefont {A.}~\bibnamefont {Gulans}},
  \bibinfo {author} {\bibfnamefont {A.~V.}\ \bibnamefont {Krasheninnikov}}, \
  and\ \bibinfo {author} {\bibfnamefont {R.~M.}\ \bibnamefont {Nieminen}},\
  }\href {\doibase 10.1103/PhysRevLett.108.235502} {\bibfield  {journal}
  {\bibinfo  {journal} {Phys. Rev. Lett.}\ }\textbf {\bibinfo {volume} {108}},\
  \bibinfo {pages} {235502} (\bibinfo {year} {2012})}\BibitemShut {NoStop}%
\bibitem [{\citenamefont {Chen}\ \emph {et~al.}(2013)\citenamefont {Chen},
  \citenamefont {Tian}, \citenamefont {Persson}, \citenamefont {Duan},\ and\
  \citenamefont {Chen}}]{Chen2013}%
  \BibitemOpen
  \bibfield  {author} {\bibinfo {author} {\bibfnamefont {X.}~\bibnamefont
  {Chen}}, \bibinfo {author} {\bibfnamefont {F.}~\bibnamefont {Tian}}, \bibinfo
  {author} {\bibfnamefont {C.}~\bibnamefont {Persson}}, \bibinfo {author}
  {\bibfnamefont {W.}~\bibnamefont {Duan}}, \ and\ \bibinfo {author}
  {\bibfnamefont {N.-X.}\ \bibnamefont {Chen}},\ }\href {\doibase
  10.1038/srep03046} {\bibfield  {journal} {\bibinfo  {journal} {Sci. Rep.}\
  }\textbf {\bibinfo {volume} {3}},\ \bibinfo {pages} {3046} (\bibinfo {year}
  {2013})}\BibitemShut {NoStop}%
\bibitem [{\citenamefont {Schabel}\ and\ \citenamefont
  {Martins}(1992)}]{Schabel1992}%
  \BibitemOpen
  \bibfield  {author} {\bibinfo {author} {\bibfnamefont {M.~C.}\ \bibnamefont
  {Schabel}}\ and\ \bibinfo {author} {\bibfnamefont {J.~L.}\ \bibnamefont
  {Martins}},\ }\href {\doibase 10.1103/PhysRevB.46.7185} {\bibfield  {journal}
  {\bibinfo  {journal} {Phys. Rev. B}\ }\textbf {\bibinfo {volume} {46}},\
  \bibinfo {pages} {7185} (\bibinfo {year} {1992})}\BibitemShut {NoStop}%
\bibitem [{\citenamefont {Zacharia}\ \emph {et~al.}(2004)\citenamefont
  {Zacharia}, \citenamefont {Ulbricht},\ and\ \citenamefont
  {Hertel}}]{Zacharia2004}%
  \BibitemOpen
  \bibfield  {author} {\bibinfo {author} {\bibfnamefont {R.}~\bibnamefont
  {Zacharia}}, \bibinfo {author} {\bibfnamefont {H.}~\bibnamefont {Ulbricht}},
  \ and\ \bibinfo {author} {\bibfnamefont {T.}~\bibnamefont {Hertel}},\ }\href
  {\doibase 10.1103/PhysRevB.69.155406} {\bibfield  {journal} {\bibinfo
  {journal} {Phys. Rev. B}\ }\textbf {\bibinfo {volume} {69}},\ \bibinfo
  {pages} {155406} (\bibinfo {year} {2004})}\BibitemShut {NoStop}%
\bibitem [{\citenamefont {Ortmann}\ \emph {et~al.}(2006)\citenamefont
  {Ortmann}, \citenamefont {Bechstedt},\ and\ \citenamefont
  {Schmidt}}]{Ortmann2006}%
  \BibitemOpen
  \bibfield  {author} {\bibinfo {author} {\bibfnamefont {F.}~\bibnamefont
  {Ortmann}}, \bibinfo {author} {\bibfnamefont {F.}~\bibnamefont {Bechstedt}},
  \ and\ \bibinfo {author} {\bibfnamefont {W.~G.}\ \bibnamefont {Schmidt}},\
  }\href {\doibase 10.1103/PhysRevB.73.205101} {\bibfield  {journal} {\bibinfo
  {journal} {Phys. Rev. B}\ }\textbf {\bibinfo {volume} {73}},\ \bibinfo
  {pages} {205101} (\bibinfo {year} {2006})}\BibitemShut {NoStop}%
\bibitem [{\citenamefont {Hasegawa}\ \emph {et~al.}(2007)\citenamefont
  {Hasegawa}, \citenamefont {Nishidate},\ and\ \citenamefont
  {Iyetomi}}]{Hasegawa2007}%
  \BibitemOpen
  \bibfield  {author} {\bibinfo {author} {\bibfnamefont {M.}~\bibnamefont
  {Hasegawa}}, \bibinfo {author} {\bibfnamefont {K.}~\bibnamefont {Nishidate}},
  \ and\ \bibinfo {author} {\bibfnamefont {H.}~\bibnamefont {Iyetomi}},\ }\href
  {\doibase 10.1103/PhysRevB.76.115424} {\bibfield  {journal} {\bibinfo
  {journal} {Phys. Rev. B}\ }\textbf {\bibinfo {volume} {76}},\ \bibinfo
  {pages} {115424} (\bibinfo {year} {2007})}\BibitemShut {NoStop}%
\bibitem [{\citenamefont {Ashton}\ \emph {et~al.}(2017)\citenamefont {Ashton},
  \citenamefont {Paul}, \citenamefont {Sinnott},\ and\ \citenamefont
  {Hennig}}]{Ashton2017}%
  \BibitemOpen
  \bibfield  {author} {\bibinfo {author} {\bibfnamefont {M.}~\bibnamefont
  {Ashton}}, \bibinfo {author} {\bibfnamefont {J.}~\bibnamefont {Paul}},
  \bibinfo {author} {\bibfnamefont {S.~B.}\ \bibnamefont {Sinnott}}, \ and\
  \bibinfo {author} {\bibfnamefont {R.~G.}\ \bibnamefont {Hennig}},\ }\href
  {\doibase 10.1103/PhysRevLett.118.106101} {\bibfield  {journal} {\bibinfo
  {journal} {Phys. Rev. Lett.}\ }\textbf {\bibinfo {volume} {118}},\ \bibinfo
  {pages} {106101} (\bibinfo {year} {2017})}\BibitemShut {NoStop}%
\bibitem [{\citenamefont {Jing}\ \emph {et~al.}(2017)\citenamefont {Jing},
  \citenamefont {Ma}, \citenamefont {Li},\ and\ \citenamefont
  {Heine}}]{Jing2017}%
  \BibitemOpen
  \bibfield  {author} {\bibinfo {author} {\bibfnamefont {Y.}~\bibnamefont
  {Jing}}, \bibinfo {author} {\bibfnamefont {Y.}~\bibnamefont {Ma}}, \bibinfo
  {author} {\bibfnamefont {Y.}~\bibnamefont {Li}}, \ and\ \bibinfo {author}
  {\bibfnamefont {T.}~\bibnamefont {Heine}},\ }\href {\doibase
  10.1021/acs.nanolett.6b05143} {\bibfield  {journal} {\bibinfo  {journal}
  {Nano Lett.}\ }\textbf {\bibinfo {volume} {17}},\ \bibinfo {pages} {1833}
  (\bibinfo {year} {2017})}\BibitemShut {NoStop}%
\bibitem [{\citenamefont {Zhao}\ \emph {et~al.}(2014)\citenamefont {Zhao},
  \citenamefont {Li},\ and\ \citenamefont {Yang}}]{Zhao2014}%
  \BibitemOpen
  \bibfield  {author} {\bibinfo {author} {\bibfnamefont {S.}~\bibnamefont
  {Zhao}}, \bibinfo {author} {\bibfnamefont {Z.}~\bibnamefont {Li}}, \ and\
  \bibinfo {author} {\bibfnamefont {J.}~\bibnamefont {Yang}},\ }\href {\doibase
  10.1021/ja5065125} {\bibfield  {journal} {\bibinfo  {journal} {J. Am. Chem.
  Soc.}\ }\textbf {\bibinfo {volume} {136}},\ \bibinfo {pages} {13313}
  (\bibinfo {year} {2014})}\BibitemShut {NoStop}%
\bibitem [{\citenamefont {Li}\ \emph {et~al.}(2016)\citenamefont {Li},
  \citenamefont {Liu}, \citenamefont {Wang},\ and\ \citenamefont
  {Li}}]{Li2016}%
  \BibitemOpen
  \bibfield  {author} {\bibinfo {author} {\bibfnamefont {F.}~\bibnamefont
  {Li}}, \bibinfo {author} {\bibfnamefont {X.}~\bibnamefont {Liu}}, \bibinfo
  {author} {\bibfnamefont {Y.}~\bibnamefont {Wang}}, \ and\ \bibinfo {author}
  {\bibfnamefont {Y.}~\bibnamefont {Li}},\ }\href@noop {} {\bibfield  {journal}
  {\bibinfo  {journal} {J. Mater. Chem. C}\ }\textbf {\bibinfo {volume} {4}},\
  \bibinfo {pages} {2155} (\bibinfo {year} {2016})}\BibitemShut {NoStop}%
\bibitem [{\citenamefont {Jiao}\ \emph {et~al.}(2015)\citenamefont {Jiao},
  \citenamefont {Ma}, \citenamefont {Gao}, \citenamefont {Bell}, \citenamefont
  {Frauenheim},\ and\ \citenamefont {Du}}]{Jiao2015}%
  \BibitemOpen
  \bibfield  {author} {\bibinfo {author} {\bibfnamefont {Y.}~\bibnamefont
  {Jiao}}, \bibinfo {author} {\bibfnamefont {F.}~\bibnamefont {Ma}}, \bibinfo
  {author} {\bibfnamefont {G.}~\bibnamefont {Gao}}, \bibinfo {author}
  {\bibfnamefont {J.}~\bibnamefont {Bell}}, \bibinfo {author} {\bibfnamefont
  {T.}~\bibnamefont {Frauenheim}}, \ and\ \bibinfo {author} {\bibfnamefont
  {A.}~\bibnamefont {Du}},\ }\href {\doibase 10.1021/acs.jpclett.5b01136}
  {\bibfield  {journal} {\bibinfo  {journal} {J. Phys. Chem. Lett.}\ }\textbf
  {\bibinfo {volume} {6}},\ \bibinfo {pages} {2682} (\bibinfo {year}
  {2015})}\BibitemShut {NoStop}%
\bibitem [{\citenamefont {Shulenburger}\ \emph {et~al.}(2015)\citenamefont
  {Shulenburger}, \citenamefont {Baczewski}, \citenamefont {Zhu}, \citenamefont
  {Guan},\ and\ \citenamefont {Tomanek}}]{Shulenburger2015}%
  \BibitemOpen
  \bibfield  {author} {\bibinfo {author} {\bibfnamefont {L.}~\bibnamefont
  {Shulenburger}}, \bibinfo {author} {\bibfnamefont {A.~D.}\ \bibnamefont
  {Baczewski}}, \bibinfo {author} {\bibfnamefont {Z.}~\bibnamefont {Zhu}},
  \bibinfo {author} {\bibfnamefont {J.}~\bibnamefont {Guan}}, \ and\ \bibinfo
  {author} {\bibfnamefont {D.}~\bibnamefont {Tomanek}},\ }\href@noop {}
  {\bibfield  {journal} {\bibinfo  {journal} {Nano Lett.}\ }\textbf {\bibinfo
  {volume} {15}},\ \bibinfo {pages} {8170} (\bibinfo {year}
  {2015})}\BibitemShut {NoStop}%
\bibitem [{\citenamefont {Sch\"utz}\ \emph {et~al.}(2017)\citenamefont
  {Sch\"utz}, \citenamefont {Maschio}, \citenamefont {Karttunen},\ and\
  \citenamefont {Usvyat}}]{Schutz2017}%
  \BibitemOpen
  \bibfield  {author} {\bibinfo {author} {\bibfnamefont {M.}~\bibnamefont
  {Sch\"utz}}, \bibinfo {author} {\bibfnamefont {L.}~\bibnamefont {Maschio}},
  \bibinfo {author} {\bibfnamefont {A.~J.}\ \bibnamefont {Karttunen}}, \ and\
  \bibinfo {author} {\bibfnamefont {D.}~\bibnamefont {Usvyat}},\ }\href
  {\doibase 10.1021/acs.jpclett.7b00253} {\bibfield  {journal} {\bibinfo
  {journal} {J. Phys. Chem. Lett.}\ }\textbf {\bibinfo {volume} {8}},\ \bibinfo
  {pages} {1290} (\bibinfo {year} {2017})}\BibitemShut {NoStop}%
\bibitem [{\citenamefont {Mostaani}\ \emph {et~al.}(2015)\citenamefont
  {Mostaani}, \citenamefont {Drummond},\ and\ \citenamefont
  {Fal'ko}}]{Mostaani2015}%
  \BibitemOpen
  \bibfield  {author} {\bibinfo {author} {\bibfnamefont {E.}~\bibnamefont
  {Mostaani}}, \bibinfo {author} {\bibfnamefont {N.~D.}\ \bibnamefont
  {Drummond}}, \ and\ \bibinfo {author} {\bibfnamefont {V.~I.}\ \bibnamefont
  {Fal'ko}},\ }\href {\doibase 10.1103/PhysRevLett.115.115501} {\bibfield
  {journal} {\bibinfo  {journal} {Phys. Rev. Lett.}\ }\textbf {\bibinfo
  {volume} {115}},\ \bibinfo {pages} {115501} (\bibinfo {year}
  {2015})}\BibitemShut {NoStop}%
\bibitem [{\citenamefont {Dubeck\'y}\ \emph {et~al.}(2016)\citenamefont
  {Dubeck\'y}, \citenamefont {Mitas},\ and\ \citenamefont
  {Jure\v{c}ka}}]{Dubecky2016}%
  \BibitemOpen
  \bibfield  {author} {\bibinfo {author} {\bibfnamefont {M.}~\bibnamefont
  {Dubeck\'y}}, \bibinfo {author} {\bibfnamefont {L.}~\bibnamefont {Mitas}}, \
  and\ \bibinfo {author} {\bibfnamefont {P.}~\bibnamefont {Jure\v{c}ka}},\
  }\href {\doibase 10.1021/acs.chemrev.5b00577} {\bibfield  {journal} {\bibinfo
   {journal} {Chem. Rev. (Washington, DC, U. S.)}\ }\textbf {\bibinfo {volume}
  {116}},\ \bibinfo {pages} {5188} (\bibinfo {year} {2016})}\BibitemShut
  {NoStop}%
\bibitem [{\citenamefont {Sansone}\ \emph {et~al.}(2016)\citenamefont
  {Sansone}, \citenamefont {Maschio}, \citenamefont {Usvyat}, \citenamefont
  {Sch\"utz},\ and\ \citenamefont {Karttunen}}]{Sansone2016}%
  \BibitemOpen
  \bibfield  {author} {\bibinfo {author} {\bibfnamefont {G.}~\bibnamefont
  {Sansone}}, \bibinfo {author} {\bibfnamefont {L.}~\bibnamefont {Maschio}},
  \bibinfo {author} {\bibfnamefont {D.}~\bibnamefont {Usvyat}}, \bibinfo
  {author} {\bibfnamefont {M.}~\bibnamefont {Sch\"utz}}, \ and\ \bibinfo
  {author} {\bibfnamefont {A.}~\bibnamefont {Karttunen}},\ }\href {\doibase
  10.1021/acs.jpclett.5b02174} {\bibfield  {journal} {\bibinfo  {journal} {J.
  Phys. Chem. Lett.}\ }\textbf {\bibinfo {volume} {7}},\ \bibinfo {pages} {131}
  (\bibinfo {year} {2016})}\BibitemShut {NoStop}%
\bibitem [{\citenamefont {Yang}\ and\ \citenamefont {Wang}(2001)}]{Yang2001}%
  \BibitemOpen
  \bibfield  {author} {\bibinfo {author} {\bibfnamefont {G.}~\bibnamefont
  {Yang}}\ and\ \bibinfo {author} {\bibfnamefont {J.}~\bibnamefont {Wang}},\
  }\href {\doibase 10.1007/s003390000537} {\bibfield  {journal} {\bibinfo
  {journal} {Appl. Phys. A: Mater. Sci. Process.}\ }\textbf {\bibinfo {volume}
  {72}},\ \bibinfo {pages} {475} (\bibinfo {year} {2001})}\BibitemShut
  {NoStop}%
\bibitem [{\citenamefont {Kresse}\ and\ \citenamefont
  {Joubert}(1999)}]{Kresse1999}%
  \BibitemOpen
  \bibfield  {author} {\bibinfo {author} {\bibfnamefont {G.}~\bibnamefont
  {Kresse}}\ and\ \bibinfo {author} {\bibfnamefont {D.}~\bibnamefont
  {Joubert}},\ }\href {\doibase 10.1103/PhysRevB.59.1758} {\bibfield  {journal}
  {\bibinfo  {journal} {Phys. Rev. B}\ }\textbf {\bibinfo {volume} {59}},\
  \bibinfo {pages} {1758} (\bibinfo {year} {1999})}\BibitemShut {NoStop}%
\bibitem [{\citenamefont {Kresse}\ and\ \citenamefont
  {Furthm{\"u}ller}(1996)}]{Kresse1996}%
  \BibitemOpen
  \bibfield  {author} {\bibinfo {author} {\bibfnamefont {G.}~\bibnamefont
  {Kresse}}\ and\ \bibinfo {author} {\bibfnamefont {J.}~\bibnamefont
  {Furthm{\"u}ller}},\ }\href {\doibase 10.1103/PhysRevB.54.11169} {\bibfield
  {journal} {\bibinfo  {journal} {Phys. Rev. B}\ }\textbf {\bibinfo {volume}
  {54}},\ \bibinfo {pages} {11169} (\bibinfo {year} {1996})}\BibitemShut
  {NoStop}%
\bibitem [{\citenamefont {Perdew}\ \emph {et~al.}(1996)\citenamefont {Perdew},
  \citenamefont {Burke},\ and\ \citenamefont {Ernzerhof}}]{Perdew1996}%
  \BibitemOpen
  \bibfield  {author} {\bibinfo {author} {\bibfnamefont {J.~P.}\ \bibnamefont
  {Perdew}}, \bibinfo {author} {\bibfnamefont {K.}~\bibnamefont {Burke}}, \
  and\ \bibinfo {author} {\bibfnamefont {M.}~\bibnamefont {Ernzerhof}},\ }\href
  {\doibase 10.1103/PhysRevLett.77.3865} {\bibfield  {journal} {\bibinfo
  {journal} {Phys. Rev. Lett.}\ }\textbf {\bibinfo {volume} {77}},\ \bibinfo
  {pages} {3865} (\bibinfo {year} {1996})}\BibitemShut {NoStop}%
\bibitem [{\citenamefont {Grimme}(2006)}]{Grimme2006}%
  \BibitemOpen
  \bibfield  {author} {\bibinfo {author} {\bibfnamefont {S.}~\bibnamefont
  {Grimme}},\ }\href {\doibase 10.1002/jcc.20495} {\bibfield  {journal}
  {\bibinfo  {journal} {J. Comput. Chem.}\ }\textbf {\bibinfo {volume} {27}},\
  \bibinfo {pages} {1787} (\bibinfo {year} {2006})}\BibitemShut {NoStop}%
\bibitem [{\citenamefont {Bachhuber}\ \emph {et~al.}(2015)\citenamefont
  {Bachhuber}, \citenamefont {von Appen}, \citenamefont {Dronskowski},
  \citenamefont {Schmidt}, \citenamefont {Nilges}, \citenamefont {Pfitzner},\
  and\ \citenamefont {Weihrich}}]{Bachhuber2015}%
  \BibitemOpen
  \bibfield  {author} {\bibinfo {author} {\bibfnamefont {F.}~\bibnamefont
  {Bachhuber}}, \bibinfo {author} {\bibfnamefont {J.}~\bibnamefont {von
  Appen}}, \bibinfo {author} {\bibfnamefont {R.}~\bibnamefont {Dronskowski}},
  \bibinfo {author} {\bibfnamefont {P.}~\bibnamefont {Schmidt}}, \bibinfo
  {author} {\bibfnamefont {T.}~\bibnamefont {Nilges}}, \bibinfo {author}
  {\bibfnamefont {A.}~\bibnamefont {Pfitzner}}, \ and\ \bibinfo {author}
  {\bibfnamefont {R.}~\bibnamefont {Weihrich}},\ }\href@noop {} {\bibfield
  {journal} {\bibinfo  {journal} {Z. Kristallogr. - Cryst. Mater.}\ }\textbf
  {\bibinfo {volume} {230}},\ \bibinfo {pages} {107} (\bibinfo {year}
  {2015})}\BibitemShut {NoStop}%
\bibitem [{Note1()}]{Note1}%
  \BibitemOpen
  \bibinfo {note} {The interlayer binding energies of graphite (21.2~meV/\r
  A$^2$) and black phosphorus (22.9~meV/\r A$^2$) obtained by neglecting the
  relaxation of the in-plane lattice parameters are in good agreement with
  those obtained in a similar way by Hanke et al.\ \cite {Hanke2011} for
  graphite, 21.8~meV/\r A$^2$, and by Sansone et al.\ \cite {Sansone2016} for
  black phosphorus, 23.0~meV/\r A$^2$.}\BibitemShut {Stop}%
\bibitem [{\citenamefont {Malliakas}\ and\ \citenamefont
  {Kanatzidis}(2013)}]{Malliakas2013}%
  \BibitemOpen
  \bibfield  {author} {\bibinfo {author} {\bibfnamefont {C.~D.}\ \bibnamefont
  {Malliakas}}\ and\ \bibinfo {author} {\bibfnamefont {M.~G.}\ \bibnamefont
  {Kanatzidis}},\ }\href {\doibase 10.1021/ja3120554} {\bibfield  {journal}
  {\bibinfo  {journal} {J. Am. Chem. Soc.}\ }\textbf {\bibinfo {volume}
  {135}},\ \bibinfo {pages} {1719} (\bibinfo {year} {2013})}\BibitemShut
  {NoStop}%
\bibitem [{\citenamefont {Langer}\ \emph {et~al.}(2014)\citenamefont {Langer},
  \citenamefont {Kisiel}, \citenamefont {Pawlak}, \citenamefont {Pellegrini},
  \citenamefont {Santoro}, \citenamefont {Buzio}, \citenamefont {Gerbi},
  \citenamefont {Balakrishnan}, \citenamefont {Baratoff}, \citenamefont
  {Tosatti},\ and\ \citenamefont {Meyer}}]{Langer2014}%
  \BibitemOpen
  \bibfield  {author} {\bibinfo {author} {\bibfnamefont {M.}~\bibnamefont
  {Langer}}, \bibinfo {author} {\bibfnamefont {M.}~\bibnamefont {Kisiel}},
  \bibinfo {author} {\bibfnamefont {R.}~\bibnamefont {Pawlak}}, \bibinfo
  {author} {\bibfnamefont {F.}~\bibnamefont {Pellegrini}}, \bibinfo {author}
  {\bibfnamefont {G.~E.}\ \bibnamefont {Santoro}}, \bibinfo {author}
  {\bibfnamefont {R.}~\bibnamefont {Buzio}}, \bibinfo {author} {\bibfnamefont
  {A.}~\bibnamefont {Gerbi}}, \bibinfo {author} {\bibfnamefont
  {G.}~\bibnamefont {Balakrishnan}}, \bibinfo {author} {\bibfnamefont
  {A.}~\bibnamefont {Baratoff}}, \bibinfo {author} {\bibfnamefont
  {E.}~\bibnamefont {Tosatti}}, \ and\ \bibinfo {author} {\bibfnamefont
  {E.}~\bibnamefont {Meyer}},\ }\href@noop {} {\bibfield  {journal} {\bibinfo
  {journal} {Nat. Mater.}\ }\textbf {\bibinfo {volume} {13}},\ \bibinfo {pages}
  {173} (\bibinfo {year} {2014})}\BibitemShut {NoStop}%
\bibitem [{\citenamefont {Ugeda}\ \emph {et~al.}(2016)\citenamefont {Ugeda},
  \citenamefont {Bradley}, \citenamefont {Zhang}, \citenamefont {Onishi},
  \citenamefont {Chen}, \citenamefont {Ruan}, \citenamefont
  {Ojeda-Aristizabal}, \citenamefont {Ryu}, \citenamefont {Edmonds},
  \citenamefont {Tsai}, \citenamefont {Riss}, \citenamefont {Mo}, \citenamefont
  {Lee}, \citenamefont {Zettl}, \citenamefont {Hussain}, \citenamefont {Shen},\
  and\ \citenamefont {Crommie}}]{Ugeda2016}%
  \BibitemOpen
  \bibfield  {author} {\bibinfo {author} {\bibfnamefont {M.~M.}\ \bibnamefont
  {Ugeda}}, \bibinfo {author} {\bibfnamefont {A.~J.}\ \bibnamefont {Bradley}},
  \bibinfo {author} {\bibfnamefont {Y.}~\bibnamefont {Zhang}}, \bibinfo
  {author} {\bibfnamefont {S.}~\bibnamefont {Onishi}}, \bibinfo {author}
  {\bibfnamefont {Y.}~\bibnamefont {Chen}}, \bibinfo {author} {\bibfnamefont
  {W.}~\bibnamefont {Ruan}}, \bibinfo {author} {\bibfnamefont {C.}~\bibnamefont
  {Ojeda-Aristizabal}}, \bibinfo {author} {\bibfnamefont {H.}~\bibnamefont
  {Ryu}}, \bibinfo {author} {\bibfnamefont {M.~T.}\ \bibnamefont {Edmonds}},
  \bibinfo {author} {\bibfnamefont {H.-Z.}\ \bibnamefont {Tsai}}, \bibinfo
  {author} {\bibfnamefont {A.}~\bibnamefont {Riss}}, \bibinfo {author}
  {\bibfnamefont {S.-K.}\ \bibnamefont {Mo}}, \bibinfo {author} {\bibfnamefont
  {D.}~\bibnamefont {Lee}}, \bibinfo {author} {\bibfnamefont {A.}~\bibnamefont
  {Zettl}}, \bibinfo {author} {\bibfnamefont {Z.}~\bibnamefont {Hussain}},
  \bibinfo {author} {\bibfnamefont {Z.-X.}\ \bibnamefont {Shen}}, \ and\
  \bibinfo {author} {\bibfnamefont {M.~F.}\ \bibnamefont {Crommie}},\
  }\href@noop {} {\bibfield  {journal} {\bibinfo  {journal} {Nat. Phys.}\
  }\textbf {\bibinfo {volume} {12}},\ \bibinfo {pages} {92} (\bibinfo {year}
  {2016})}\BibitemShut {NoStop}%
\bibitem [{\citenamefont {Silva-Guill{\'e}n}\ \emph {et~al.}(2016)\citenamefont
  {Silva-Guill{\'e}n}, \citenamefont {Ordej{\'o}n}, \citenamefont {Guinea},\
  and\ \citenamefont {Canadell}}]{Silva2016}%
  \BibitemOpen
  \bibfield  {author} {\bibinfo {author} {\bibfnamefont {J.~{\'A}.}\
  \bibnamefont {Silva-Guill{\'e}n}}, \bibinfo {author} {\bibfnamefont
  {P.}~\bibnamefont {Ordej{\'o}n}}, \bibinfo {author} {\bibfnamefont
  {F.}~\bibnamefont {Guinea}}, \ and\ \bibinfo {author} {\bibfnamefont
  {E.}~\bibnamefont {Canadell}},\ }\href@noop {} {\bibfield  {journal}
  {\bibinfo  {journal} {2D Mater.}\ }\textbf {\bibinfo {volume} {3}},\ \bibinfo
  {pages} {035028} (\bibinfo {year} {2016})}\BibitemShut {NoStop}%
\bibitem [{\citenamefont {Booth}\ \emph {et~al.}(2013)\citenamefont {Booth},
  \citenamefont {Gr{\"u}neis}, \citenamefont {Kresse},\ and\ \citenamefont
  {Alavi}}]{Booth2013}%
  \BibitemOpen
  \bibfield  {author} {\bibinfo {author} {\bibfnamefont {G.~H.}\ \bibnamefont
  {Booth}}, \bibinfo {author} {\bibfnamefont {A.}~\bibnamefont {Gr{\"u}neis}},
  \bibinfo {author} {\bibfnamefont {G.}~\bibnamefont {Kresse}}, \ and\ \bibinfo
  {author} {\bibfnamefont {A.}~\bibnamefont {Alavi}},\ }\href@noop {}
  {\bibfield  {journal} {\bibinfo  {journal} {Nature}\ }\textbf {\bibinfo
  {volume} {493}},\ \bibinfo {pages} {365} (\bibinfo {year}
  {2013})}\BibitemShut {NoStop}%
\bibitem [{\citenamefont {Whittaker}(1982)}]{Whittaker1982}%
  \BibitemOpen
  \bibfield  {author} {\bibinfo {author} {\bibfnamefont {E.~J.~W.}\
  \bibnamefont {Whittaker}},\ }\href@noop {} {\bibfield  {journal} {\bibinfo
  {journal} {Mineral. Mag.}\ }\textbf {\bibinfo {volume} {46}},\ \bibinfo
  {pages} {398} (\bibinfo {year} {1982})}\BibitemShut {NoStop}%
\end{thebibliography}%

\end{document}